\documentclass[aps,preprint,nofootinbib,showpacs,superscriptaddress]{revtex4}
\usepackage{amssymb}
\usepackage{amsmath}
\usepackage{graphicx}
\usepackage{multirow}

\begin{document}

\title{Next-to-leading order QCD predictions for $A^{0}\gamma$ associated production at the CERN Large Hadron Collider}
\author{Liang Dai}
\affiliation{Department of Physics and State Key Laboratory of
Nuclear Physics and Technology, Peking University, Beijing 100871,
China}
\author{Ding Yu Shao}
\affiliation{Department of Physics and State Key Laboratory of
Nuclear Physics and Technology, Peking University, Beijing 100871,
China}
\author{Chong Sheng Li}
\email{csli@pku.edu.cn}
\affiliation{Department of Physics and State Key Laboratory of
Nuclear Physics and Technology, Peking University, Beijing 100871,
China}
\affiliation{Center for High Energy Physics, Peking University, Beijing 100871, China}
\author{Jun Gao}
\affiliation{Department of Physics and State Key Laboratory of
Nuclear Physics and Technology, Peking University, Beijing 100871,
China}
\author{Hao Zhang}
\affiliation{Department of Physics and State Key Laboratory of
Nuclear Physics and Technology, Peking University, Beijing 100871,
China}

\date{\today}

\begin{abstract}
 We calculate the complete next-to-leading-order (NLO) QCD corrections (including SUSY QCD corrections) to the inclusive total cross sections of the associated production processes $pp\rightarrow A^{0}\gamma+X$ in the minimal supersymmetric standard model (MSSM) at the CERN Large Hadron Collider (LHC). Our results show that the enhancement of the total cross sections from the NLO QCD corrections can reach $25\%\sim15\%$ for $200$~GeV$<m_{A}<500$~GeV and $\tan\beta=50$. The scale dependence of the total cross section is improved by the NLO corrections in general. We also show the Monte Carlo simulation results for the $\tau^{+}\tau^{-}+\gamma$ signature including the complete NLO QCD effects, and find an observable signature above the standard model (SM) background for a normal luminosity of $100$~fb$^{-1}$ at the LHC.

\end{abstract}


\maketitle

\section{INTRODUCTION}

Electroweak symmetry breaking (EWSB) plays a key role in the current research of elementary particles.
However, the experimental effort to validate
the Higgs mechanism on the CERN Large Hadron Collider (LHC) with a center of mass energy
$\sqrt{s}=14$~TeV and a luminosity of $100$~fb$^{-1}$ per year~\cite{LHC}, is a great challenge. In the standard model (SM) of particle
physics, there is only one Higgs particle, which is expected to be lurking somewhere close to the
experimental lower bound of $114.4$~GeV set by LEP2~\cite{LEP2-bound}.
In the minimal supersymmetric standard model (MSSM), two complex Higgs doublets are introduced
to eliminate gauge anomaly~\cite{susy-primer}, resulting in
two CP-even ($h^{0},H^{0}$) and one CP-odd ($A^{0}$) neutral Higgs
bosons, as well as a pair of charged Higgs bosons. The Higgs sector
of the MSSM, at leading order, is characterized by two parameters: one
is $m_{A}$ the mass of the pseudo-scalar Higgs boson, and the other
$\tan\beta$ the ratio of up- and down-Higgs doublet vacuum expectation value (VEV). Particularly,
current experiments hint a scenario with large $\tan\beta\gtrsim45$
and thus large couplings between the pseudo-scalar Higgs and down-type
quarks~\cite{large-tb}.

At the LHC neutral Higgs bosons are mainly produced via gluon-gluon
fusion channel $gg\rightarrow\phi$ ~\cite{gg-fusion1,gg-fusion2,gg-fusion3,gg-fusion4,gg-fusion5,gg-fusion6,gg-fusion7,gg-fusion8,gg-fusion9,gg-fusion10}.
The weak boson fusion channel $qq\rightarrow qqV^{*}V^{*}\rightarrow qqh^{0}/qqH^{0}$
~\cite{weak-boson-fusion1,weak-boson-fusion2,weak-boson-fusion3} as
well as the associated production channel with weak bosons~\cite{aso-weak-boson1,aso-weak-boson2,aso-weak-boson3}
also have significant contributions. Other production channels
also have been studied, such as Higgs boson pair production~\cite{pair-prod1,pair-prod2,pair-prod3,pair-prod4}
and associated production with top quark pair~\cite{aso-tt-pair-prod1,aso-tt-pair-prod2,aso-tt-pair-prod3,aso-tt-pair-prod4}.
Nevertheless, the identification of the Higgs signature is difficult
due to large QCD backgrounds against various Higgs particle decay
modes. Recently, the Higgs boson and photon associated production channel has aroused interest~\cite{tree-level}.
For neutral Higgs boson and photon associated production, the otherwise
dominating gluon fusion channel is forbidden via C-parity conservation, so
quark-antiquark annihilation becomes dominant. In the case
of CP-odd Higgs $A^{0}$ produced with a photon in a large $\tan\beta$
MSSM scenario, bottom quark annihilation $b\bar{b}\rightarrow A^{0}\gamma$
is of particular importance due to the large Yukawa coupling enhanced
by the large $\tan\beta$. That compensates for the relatively small parton density
of the bottom quark and the suppression from the QED
vertex. Besides, associated production arising from weak boson fusion has
also been studied in Ref.~\cite{WW-fusion}.

Since the bottom quark initial state contribution to $A^{0}\gamma$ associated production is sensitive to the bottom quark Yukawa coupling, the measurement of this channel at the LHC can give detailed information of the Higgs coupling to the bottom quark. To provide a precise prediction of this associated production channel, we calculate the NLO QCD corrections to the total cross section and the kinematic distributions. In addition
to effects from virtual or real gluons, loop effects from massive supersymmetry particles (the SUSY QCD effects), such as the sbottoms and the gluino, are also considered. Dimensional regularization scheme
(DREG) (with naive $\gamma_{5}$~\cite{DREG}) is adopted to regularize both ultraviolet (UV) and infrared (IR) divergences, which is equivalent to conventional supersymmetry-preserving dimensional reduction
scheme (DRED) at the NLO level~\cite{DRED,DRED=DREG}. For simplicity, we neglect the bottom quark mass except for in the Yukawa coupling. According to the simplified Aivazis-Collins-Olness-Tung scheme~\cite{ACOT1,ACOT2,ACOT3}, such approximation is justified if the bottom quark appears as an initial parton.

The paper is organized as follows. In Sec.~\ref{sec:LO}, brief results for leading-order
(LO) calculations are presented. In Sec.~\ref{sec:NLO}, we present detailed calculations of NLO QCD corrections. In Sec.~\ref{sec:SIMULATION}, we discuss a Monte Carlo simulation of the Higgs signature from the decay mode $A^{0}\rightarrow\tau^{+}\tau^{-}$. In Sec.~\ref{sec:NUMERICAL}, we provide numerical
results for the total cross section and the differential cross sections with varying model parameters. Monte Carlo simulation results are also shown there.

\section{LEADING-ORDER CROSS SECTION FOR NEUTRAL HIGGS AND PHOTON ASSOCIATED
PRODUCTION\label{sec:LO} }

\begin{figure}[H]
\begin{centering}
\includegraphics[scale=0.5]{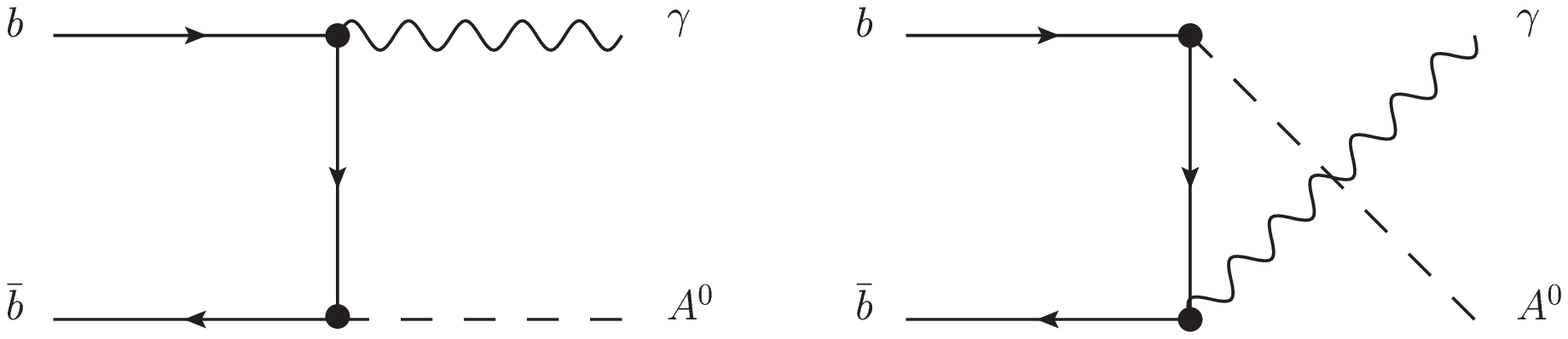}
\par\end{centering}

\caption{\label{fig:LO feynman diagrams}Tree level Feynman diagrams for
$b\bar{b}\rightarrow\gamma A^{0}$}

\end{figure}

The LO cross section for $pp\rightarrow\gamma A^{0}$ in the MSSM
has been studied in Ref.~\cite{tree-level}. At tree level the only partonic
subprocess is $b(p_{1})\bar{b}(p_{2})\rightarrow\gamma(p_{3})A^{0}(p_{4})$,
and the corresponding two Feynman diagrams are shown in Fig.~\ref{fig:LO feynman diagrams}.
The gluon-gluon fusion channel $gg\rightarrow\gamma A^{0}$ is forbidden by C-parity conservation~\cite{C-parity1,C-parity2,C-parity3}. In the tree level result we keep
a finite bottom quark mass denoted as $m_{b}$. The cross section can be written as
\begin{equation}
d\hat{\sigma}^{LO}=\frac{1}{2\Phi}dPS^{(2)}\overline{|\mathcal{M}^{B}|^{2}},
\end{equation}
where $dPS^{(2)}$ is the 2-body final-state phase space and $1/2\Phi$ is
the flux factor. The explicit expression for the differential cross
section after averaging over spins and colors can be written as
\begin{equation}
\frac{d\hat{\sigma}}{d\hat{t}}=\frac{\alpha_{em}Q_{b}^{2}\lambda_{\phi}^{2}}{4N_{c}(1-4r_{b})}\Big\{\frac{F_{1}^{\phi}(\hat{s})}{(\hat{t}-m_{b}^{2})(\hat{u}-m_{b}^{2})}+F_{2}^{\phi}(\hat{s})\Big[\frac{1}{(\hat{t}-m_{b}^{2})^{2}}+\frac{1}{(\hat{u}-m_{b}^{2})^{2}}\Big]\Big\},
\end{equation}
with
\begin{equation}
F_{1}^{\phi}(\hat{s})=(1-r_{\phi})^{2}+2r_{\phi}(1-2r_{b}),\qquad F_{2}^{\phi}(\hat{s})=-2r_{b}r_{\phi},
\end{equation}
where $Q_{b}=-1/3$ is the electric charge quantum number of the bottom quark, $N_{c}$ the number
of quark color, $r_{b}=m_{b}^{2}/\hat{s},$$r_{\phi}=m_{A}^{2}/\hat{s}$,
and $\lambda_{\phi}=-i\frac{m_{b}}{v}\tan\beta$ is the Yukawa coupling
in MSSM which is proportional to the bottom quark mass. Here $v=2m_{W}/g$
is the SM Higgs field VEV. In addition, the Mandelstam variables
for $2\rightarrow2$ scattering process are introduced
\begin{equation}
\hat{s}=(p_{1}+p_{2})^{2},\qquad\hat{t}=(p_{1}-p_{3})^{2},\qquad\hat{u}=(p_{1}-p_{4})^{2}.
\end{equation}
The hadronic cross section for $pp\rightarrow\gamma A^{0}$ at the LO
is obtained straightforwardly by convoluting the parton level cross
section with the parton distribution function (PDF),
\begin{equation}
\sigma^{B}=\int dx_{1}dx_{2}\big[G_{b/p}(x_{1},\mu_{F})G_{\bar{b}/p}(x_{2},\mu_{F})+(x_{1}\leftrightarrow x_{2})\big]\hat{\sigma}^{B},
\end{equation}
where $\mu_{F}$ is the factorization scale.

\section{NLO QCD calculations\label{sec:NLO}}

The NLO QCD correction to $\gamma A^{0}$ associated production consists of two parts. The virtual corrections account for virtual gluons as well as virtual supersymmetric particles such as the gluino $\tilde{g}$ and the sbottoms $\tilde{b}_{1,2}$ in the loop diagrams. The real corrections result from the radiation of a real gluon or a massless bottom (anti-)quark. For the NLO calculations we follow the
convention to work in $D=4-2\epsilon$ dimensions and adopt the dimensional
regularization approach (DREG) to regulate both the ultroviolet (UV)
and the infrared (IR) divergences. As a good approximation, we take the
bottom (anti-)quark mass to be zero except for in the Yukawa coupling.

\subsection{Virtual corrections}

\begin{figure}[H]
\begin{centering}
\includegraphics[scale=0.5]{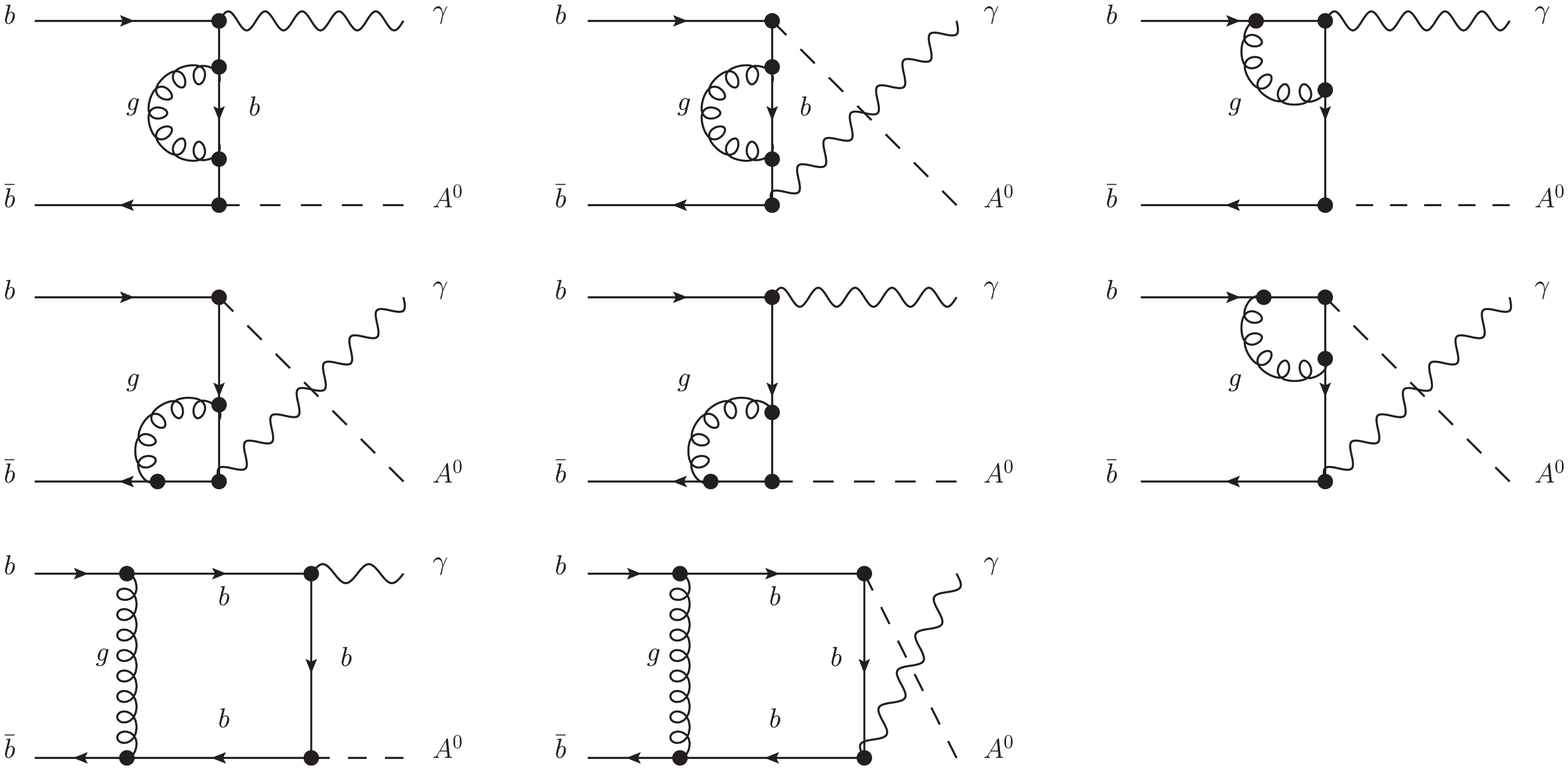}
\par\end{centering}

\caption{\label{fig:SM-virtual}The loop diagrams related to virtual
gluon: propagator, vertex and box diagram corrections}

\end{figure}

\begin{figure}[H]
\begin{centering}
\includegraphics[scale=0.5]{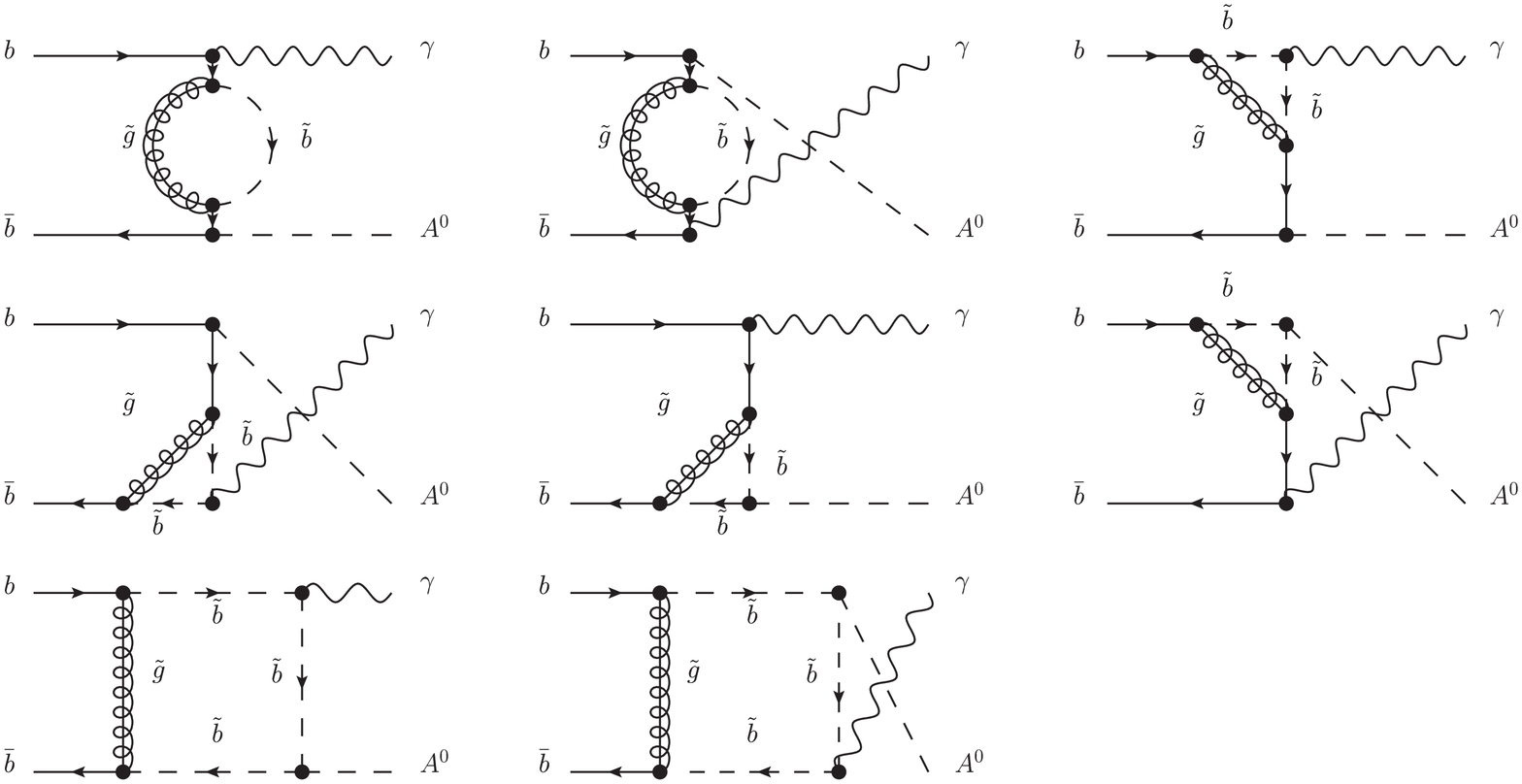}
\par\end{centering}

\caption{\label{fig:SUSY-virtual}The loop diagrams related to virtual
gluino and sbottoms: propagator, vertex and box diagram corrections}

\end{figure}

The one-loop virtual corrections involve both the SM QCD contribution
(8 diagrams as shown in Fig.\ref{fig:SM-virtual}) and the SUSY QCD contribution (another 8 diagrams as shown in Fig.\ref{fig:SUSY-virtual}). Either part is UV divergent. For the gluon
loops we adopt $\overline{MS}$ renormalization scheme to absorb those
infinities, while for the SUSY particle loops we use the on-shell
renormalization scheme instead.

\begin{figure}[H]
\begin{centering}
\includegraphics[scale=0.5]{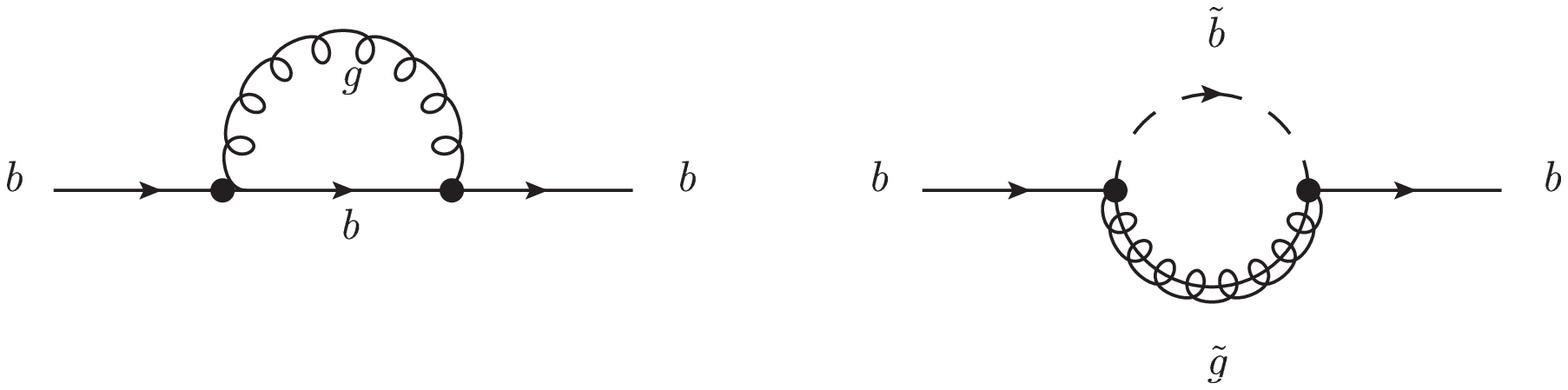}
\par\end{centering}

\caption{\label{fig:One-loop-QCD-propagator}Self-energy diagrams
for the bottom quark}

\end{figure}

\begin{figure}[H]
\begin{centering}
\includegraphics[scale=0.5]{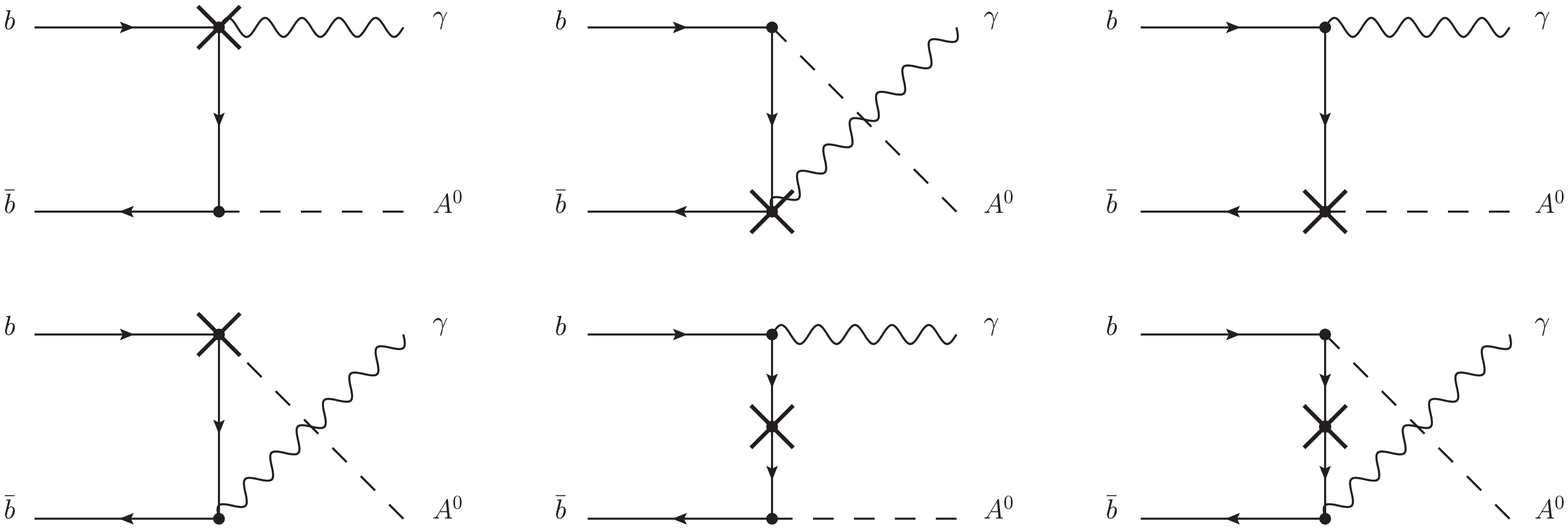}
\par\end{centering}

\caption{\label{fig:CT-diagrams}Couter-term diagrams: wavefunction,
mass and vertex renormalization}

\end{figure}

The virtual correction is given by interfering the one-loop amplitude with
the Born amplitude
\begin{equation}
d\hat{\sigma}^{V}=\frac{1}{2\hat{s}}dPS^{(2)}2Re(\overline{\mathcal{M}^{V}\cdot\mathcal{M}^{B}}),
\end{equation}
where $dPS^{(2)}$ is the 2-body final-state phase space and the flux
factor is reduced to $1/2\hat{s}$ for massless (anti-)quark. In order
to absorb all UV divergences, we introduce the renormalized bottom quark wavefunction
for both the left-handed and the right-handed components $\psi_{bL,R}$ and
the renormalized mass $m_{b}$, which are related to the bare mass $m_{b0}$ and the bare wavefunction $\psi_{b0}$ by
\begin{align}
m_{b0} & =m_{b}+\delta m_{b},\nonumber\\
\psi_{b0} & =(1+\delta Z_{bL})^{1/2}\psi_{bL}+(1+\delta Z_{bR})^{1/2}\psi_{bR},
\end{align}
with $\psi_{bL,R}=(1\mp\gamma_{5})\psi_{b}/2$. By calculating the self-energy
diagrams of the bottom quark propagator (shown in Fig.\ref{fig:One-loop-QCD-propagator}),
we obtain explicit expressions for the counter-terms which are in accordance
with the results in Ref.~\cite{JIN,ZHU},
\begin{align}
\Big(\frac{\delta m_{b}}{m_{b}}\Big)_{SM} & =-\frac{\alpha_{s}}{4\pi}C_{F}C(\epsilon)\frac{3}{\epsilon_{UV}},\nonumber\\
(\delta Z_{bL})_{SM} & =(\delta Z_{bR})_{SM}=\frac{\alpha_{s}}{4\pi}C_{F}C(\epsilon)\Big\{-\frac{3}{\epsilon_{UV}}+\frac{3}{\epsilon_{IR}}\Big\},\nonumber\\
\Big(\frac{\delta m_{b}}{m_{b}}\Big)_{SUSY}&= -\frac{\alpha_{s}}{4\pi}C_{F}\underset{i=1,2}{\sum}\big\{B_{1}(0,m_{\tilde{g}}^{2},m_{\tilde{b}_{i}}^{2})-\frac{m_{\tilde{g}}}{m_{b}}\sin2\theta_{\tilde{b}}(-1)^{i}B_{0}(0,m_{\tilde{g}}^{2},m_{\tilde{b}_{i}}^{2})\big\},\\
(\delta Z_{bL})_{SUSY} & =\frac{\alpha_{s}}{2\pi}C_{F}\underset{i=1,2}{\sum}(R_{i1}^{\tilde{b}})^{2}B_{1}(0,m_{\tilde{g}}^{2},m_{\tilde{b}_{i}}^{2}),\nonumber\\
(\delta Z_{bR})_{SUSY} & =\frac{\alpha_{s}}{2\pi}C_{F}\underset{i=1,2}{\sum}(R_{i2}^{\tilde{b}})^{2}B_{1}(0,m_{\tilde{g}}^{2},m_{\tilde{b}_{i}}^{2}),\nonumber
\end{align}
where $C_{F}=4/3$, $C(\epsilon)=\frac{\Gamma(1-\epsilon)}{\Gamma(1-2\epsilon)}(\frac{4\pi\mu_{R}^{2}}{\hat{s}})^{\epsilon}$
and $B_{0,1}$ are the two-point integrals~\cite{scalar-integrals}
, as listed explicitly below
\begin{align}
B_{0}(0,m_{1}^{2},m_{2}^{2}) & =C(\epsilon)\Big\{\frac{1}{\epsilon_{UV}}-\frac{m_{1}^{2}\ln\frac{m_{1}^{2}}{\hat{s}}-m_{2}^{2}\ln\frac{m_{2}^{2}}{\hat{s}}}{m_{1}^{2}-m_{2}^{2}}+1\Big\},\nonumber\\
B_{1}(0,m_{1}^{2},m_{2}^{2}) & =C(\epsilon)\Big\{-\frac{1}{2\epsilon_{UV}}+\frac{2m_{1}^{4}\ln\frac{m_{1}^{2}}{\hat{s}}-3m_{1}^{4}+4m_{1}^{2}m_{2}^{2}-m_{2}^{4}+2m_{2}^{2}(m_{2}^{2}-2m_{1}^{2})\ln\frac{m_{2}^{2}}{\hat{s}}}{4(m_{1}^{2}-m_{2}^{2})^{2}}\Big\},
\end{align}
where $m_{\tilde{b}_{1,2}}$ are the sbottom masses, $m_{\tilde{g}}$ is
the gluino mass, and $R^{\tilde{b}}$ is a $2\times2$ rotation matrix which transforms
the gauge eigenstates into the mass eigenstates,
\begin{equation}
\left(\begin{array}{c}
\tilde{b}_{1}\\
\tilde{b}_{2}\end{array}\right)=R^{\tilde{b}}\left(\begin{array}{c}
\tilde{b}_{L}\\
\tilde{b}_{R}\end{array}\right),\qquad R^{\tilde{b}}=\left(\begin{array}{cc}
\cos\theta_{\tilde{b}} & \sin\theta_{\tilde{b}}\\
-\sin\theta_{\tilde{b}} & \cos\theta_{\tilde{b}}\end{array}\right),
\end{equation}
with $0\leq\theta_{\tilde{b}}\leq\pi$ by convention. Furthermore,
the sbottom mass eigenvalues are solved by diagonalizing $M_{\tilde{b}}^2$,
\begin{equation}
\left(\begin{array}{cc}
m_{\tilde{b}_{1}}^{2} & 0\\
0 & m_{\tilde{b}_{2}}^{2}\end{array}\right)=R^{\tilde{b}}M_{\tilde{b}}^{2}(R^{\tilde{b}})^{\dagger},\qquad M_{\tilde{b}}^{2}=\left(\begin{array}{cc}
m_{\tilde{b}_{L}}^{2} & a_{b}m_{b}\\
a_{b}m_{b} & m_{\tilde{b}_{R}}^{2}\end{array}\right),
\end{equation}
with
\begin{gather}
m_{\tilde{b}_{L}}^{2}=M_{\tilde{Q}}^{2}+m_{b}^{2}+m_{Z}^{2}\cos2\beta C_{bL},\nonumber\\
m_{\tilde{b}_{R}}^{2}=M_{\tilde{D}}^{2}+m_{b}^{2}-m_{Z}^{2}\cos2\beta C_{bR},\nonumber\\
a_{b}=A_{b}-\mu\tan\beta.
\end{gather}
Here $C_{bL}=-1/2+\sin^{2}\theta_{W}/3$, $C_{bR}=\sin^{2}\theta_{W}/3$,
and $M_{\tilde{b}}^{2}$ is the sbottom mass matrix. $M_{\tilde{Q},\tilde{D}}^{2}$
and $A_{b}$ are soft SUSY-breaking parameters, and $\mu$ is the Higgsino
mass parameter. Since the Yukawa coupling is proportional to the bottom quark
mass, the renormalized vertex is obtained by expressing the bare mass $m_{b0}$ in terms of the renormalized mass $m_{b}$ plus a counter term $\delta m_{b}$,
\begin{equation}
-i\frac{m_{b0}}{v}\tan\beta=-i\Big[1+\Big(\frac{\delta m_{b}}{m_{b}}\Big)_{SM}+\Big(\frac{\delta m_{b}}{m_{b}}\Big)_{SUSY}\Big]\frac{m_{b}}{v}\tan\beta.
\end{equation}
All the counter-term diagrams are shown in Fig.\ref{fig:CT-diagrams}.

With the one-loop counter-terms we write the renormalized virtual amplitude
as
\begin{equation}
\mathcal{M}^{V}=\mathcal{M}_{SM}^{unren}+\mathcal{M}_{SUSY}^{unren}+\mathcal{M}^{con}.
\end{equation}
The details of the calculation include the traditional
Passarino-Veltman reduction procedure, in which Feynman amplitudes
are reduced to master scalar integrals~\cite{Passarino-Veltman}. Here we list the analytic results for all the divergent scalar integrals (only the real part is kept) involved in the calculation,
\begin{align}
B_{0}(\hat{t},0,0)=&C(\epsilon)\Big[\frac{1}{\epsilon_{UV}}+2-\ln\Big(\frac{-\hat{t}}{\hat{s}}\Big)\Big],\nonumber\\
B_{0}(\hat{u},0,0)=&C(\epsilon)\Big[\frac{1}{\epsilon_{UV}}+2-\ln\Big(\frac{-\hat{u}}{\hat{s}}\Big)\Big],\nonumber\\
B_{0}(m_{A}^{2},0,0)=&C(\epsilon)\Big[\frac{1}{\epsilon_{UV}}+2-\ln r_{\phi}\Big],\nonumber\\
C_{0}(0,0,\hat{t},0,0,0)=&C(\epsilon)\Big(\frac{1}{-\hat{t}}\Big)\Big[-\frac{1}{\epsilon_{IR}^{2}}+\frac{\ln(-\hat{t}/\hat{s})}{\epsilon_{IR}}-\frac{1}{2}\ln^{2}\Big(\frac{-\hat{t}}{\hat{s}}\Big)-\frac{\pi^{2}}{6}\Big],\nonumber\\
C_{0}(0,0,\hat{u},0,0,0)=&C(\epsilon)\Big(\frac{1}{-\hat{u}}\Big)\Big[-\frac{1}{\epsilon_{IR}^{2}}+\frac{\ln(-\hat{u}/\hat{s})}{\epsilon_{IR}}-\frac{1}{2}\ln^{2}\Big(\frac{-\hat{u}}{\hat{s}}\Big)-\frac{\pi^{2}}{6}\Big],\nonumber\\
C_{0}(0,m_{A}^{2},\hat{s},0,0,0)=&C(\epsilon)\Big(\frac{1}{1-r_{\phi}}\Big)\Big[\frac{\ln r_{\phi}}{\epsilon_{IR}}-\frac{\ln^{2}r_{\phi}}{2}\Big],\nonumber\\
C_{0}(0,m_{A}^{2},\hat{t},0,0,0)=&C(\epsilon)\Big(\frac{1}{m_{A}^{2}-\hat{t}}\Big)\Big[\frac{1}{\epsilon_{IR}}\ln\Big(\frac{-\hat{t}}{m_{A}^{2}}\Big)-\frac{1}{2}\ln r_{\phi}\ln\Big(\frac{-\hat{t}}{m_{A}^{2}}\Big)\nonumber\\
&-\frac{1}{2}\ln\Big(\frac{-\hat{t}}{\hat{s}}\big)\ln\Big(\frac{-\hat{t}}{m_{A}^{2}}\Big)-\frac{\pi^{2}}{2}\Big],\nonumber\\
C_{0}(0,m_{A}^{2},\hat{u},0,0,0)=&C(\epsilon)\Big(\frac{1}{m_{A}^{2}-\hat{u}}\Big)\Big[\frac{1}{\epsilon_{IR}}\ln\Big(\frac{-\hat{u}}{m_{A}^{2}}\Big)-\frac{1}{2}\ln r_{\phi}\ln\Big(\frac{-\hat{u}}{m_{A}^{2}}\Big)\\
&-\frac{1}{2}\ln\Big(\frac{-\hat{u}}{\hat{s}}\big)\ln\Big(\frac{-\hat{u}}{m_{A}^{2}}\Big)-\frac{\pi^{2}}{2}\Big],\nonumber\\
D_{0}(0,m_{A}^{2},0,0,\hat{t},\hat{s},0,0,0,0)=&C(\epsilon)\Big(\frac{1}{-\hat{t}}\Big)\Big[-\frac{2}{\epsilon_{IR}^{2}}+\frac{2}{\epsilon_{IR}}\ln\Big(\frac{-\hat{t}}{m_{A}^{2}}\Big)+\ln r_{\phi}^{2}\nonumber\\
&+2\Big({\rm{Li}}_{2}\Big(\frac{\hat{s}-m_{A}^{2}}{\hat{s}}\Big)-{\rm{Li}}_{2}\Big(\frac{\hat{s}-m_{A}^{2}}{\hat{t}}\Big)\Big)-\pi^{2}\Big],\nonumber\\
D_{0}(0,m_{A}^{2},0,0,\hat{u},\hat{s},0,0,0,0)=&C(\epsilon)\Big(\frac{1}{-\hat{u}}\Big)\Big[-\frac{2}{\epsilon_{IR}^{2}}+\frac{2}{\epsilon_{IR}}\ln\Big(\frac{-\hat{u}}{m_{A}^{2}}\Big)+\ln r_{\phi}^{2}\nonumber\\
&+2\Big({\rm{Li}}_{2}\Big(\frac{\hat{s}-m_{A}^{2}}{\hat{s}}\Big)-{\rm{Li}}_{2}\Big(\frac{\hat{s}-m_{A}^{2}}{\hat{u}}\Big)\Big)-\pi^{2}\Big],\nonumber
\end{align}
We then find that the renormalized amplitude $\mathcal{M}^{V}$ is UV finite, but still
contains IR poles, which is given by
\begin{equation}
\mathcal{M}^{V}=\frac{\alpha_{s}}{2\pi}C(\epsilon)\Big\{\frac{A_{2}^{V}}{\epsilon_{IR}^{2}}+\frac{A_{1}^{V}}{\epsilon_{IR}}\Big\}\mathcal{M}^{B},
\end{equation}
with
\begin{equation}
A_{2}^{V}=-C_{F},\qquad A_{1}^{V}=-\frac{3}{2}C_{F},
\end{equation}
which demonstrates that the IR divergent part is factorized and consists
of both soft and collinear singularities. The former is canceled
when we combine the virtual corrections with the real corrections, while
the latter can be canceled by adopting the mass factorization procedure.

\subsection{Real gluon emission}

\begin{figure}[H]
\begin{centering}
\includegraphics[scale=0.5]{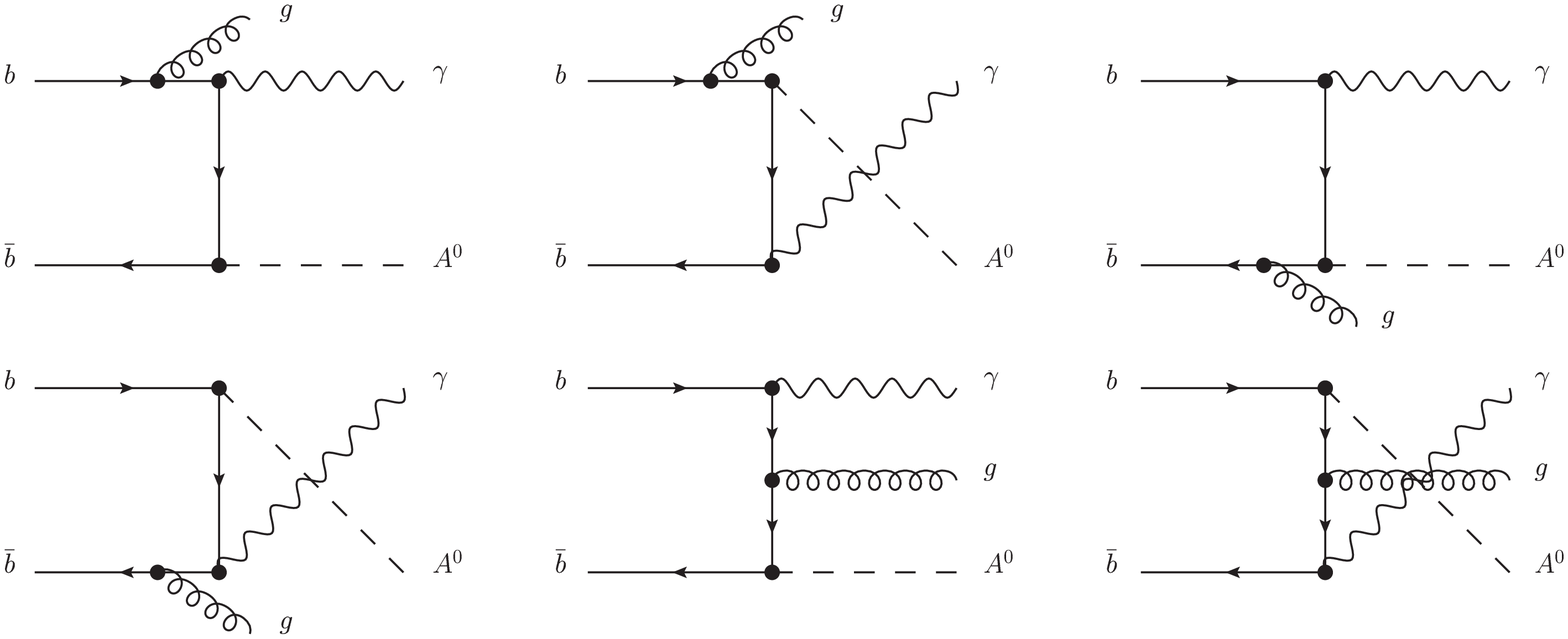}
\par\end{centering}

\caption{\label{fig:real-emission}Feynman diagrams for real gluon emission}

\end{figure}

The Feynman diagrams for the radiation of a real gluon $b(p_{1})\bar{b}(p_{2})\rightarrow\gamma(p_{3})A^{0}(p_{4})g(p_{5})$
are shown in Fig.\ref{fig:real-emission}. The partonic
cross section can be written as
\begin{equation}
d\hat{\sigma}^{R}=\frac{1}{2\hat{s}}dPS^{(3)}\overline{|\mathcal{M}^{B}|^{2}},
\end{equation}

The 3-body phase space integration for real gluon emission contains soft and collinear singularities. We adopt the two cutoff phase space slicing method~\cite{PS-slicing} to isolate all the
IR singularities, which introduces two small cutoffs $\delta_{s}$ and $\delta_{c}$
to divide the phase space into three parts.

First, the soft cutoff $\delta_{s}$ separates the phase space into
the soft region $E_{5}\leq\delta_{s}\sqrt{\hat{s}}/2$ and the hard region otherwise in the partonic center of mass (CM) frame. Thus the partonic cross section can be written as a sum of the contributions from both regions,
\begin{equation}
\hat{\sigma}^{R}=\hat{\sigma}^{S}+\hat{\sigma}^{H}.
\end{equation}
Furthermore, the hard piece can be divided into two sub-regions by
introducing a collinear cutoff $\delta_{c}$. Within the hard collinear
region $(p_{1}+p_{5})^{2}\leq\delta_{c}\hat{s}$ or $(p_{2}+p_{5})^{2}\leq\delta_{c}\hat{s}$
all the collinear divergences are isolated, leaving the hard non-collinear
region free of any IR singularities. Similarly we have for the partonic cross section
\begin{equation}
\hat{\sigma}^{H}=\hat{\sigma}^{HC}+\hat{\sigma}^{\overline{HC}}.
\end{equation}

Below we proceed to discuss the details of calculation in each region
of the phase space.

\subsubsection{Hard non-collinear region}

For the hard non-collinear region
where no IR singularity is present, the phase space integration can be calculated numerically. For
the 3-body phase space a convenient parameterization with 4 non-trivial parameters is given below,
\begin{equation}
dPS^{(3)}=\frac{\hat{s}}{32(2\pi)^{4}}dX_{1}dX_{2}d\cos\theta d\varphi.
\end{equation}
Here $-1\leq\cos\theta\leq1$ and $0\leq\varphi\leq2\pi$ represent
the solid angle in the CM frame into which the final-state photon is scattered. Besides,
$X_{1,2}$ are dimensionless variables which determine the final state
energy in the partonic CM frame through
\begin{equation}
E_{3}=\frac{\sqrt{\hat{s}}}{2}(1-X_{2}),\qquad E_{4}=\frac{\sqrt{\hat{s}}}{2}(X_{1}+X_{2}),\qquad E_{5}=\frac{\sqrt{\hat{s}}}{2}(1-X_{1}).
\end{equation}
The integration region for them is inside the unit square in
the parameter plane and are subject to kinematic constraints $X_{1}+X_{2}\leq1+r_{\phi}$
and $X_{1}X_{2}\geq r_{\phi}$.

\subsubsection{Soft region}

In the limit of vanishing gluon energy (the eikonal approximation), the squared matrix element for real
gluon emission can be factorized into the Born piece multiplied by
an eikonal factor $\Phi_{eik}$
\begin{equation}
\overline{|\mathcal{M}^{R}(b\bar{b}\rightarrow\gamma A^{0}+g)|^{2}}\overset{soft}{\longrightarrow}(4\pi\alpha_{s}\mu_{R}^{2\epsilon})\overline{|\mathcal{M}^{B}|^{2}}\Phi_{eik},
\end{equation}
where the eikonal factor can be written explicitly
\begin{equation}
\Phi_{eik}=C_{F}\Big\{\frac{-p_{1}^{2}}{(p_{1}\cdot p_{5})^{2}}+\frac{-p_{2}^{2}}{(p_{2}\cdot p_{5})^{2}}+\frac{2(p_{1}\cdot p_{2})}{(p_{1}\cdot p_{5})(p_{2}\cdot p_{5})}\Big\}=C_{F}\frac{\hat{s}}{(p_{1}\cdot p_{5})(p_{2}\cdot p_{5})}.
\end{equation}
Meanwhile the 3-body phase space is factorized into the following
form
\begin{equation}
dPS^{(3)}(b\bar{b}\rightarrow\gamma A^{0}+g)\overset{soft}{\longrightarrow}dPS^{(2)}(b\bar{b}\rightarrow\gamma A^{0})dS,
\end{equation}
with $dS$ the soft gluon phase space to be integrated
\begin{equation}
dS=\frac{1}{\pi}(\frac{4}{\hat{s}})^{-\epsilon}\int_{0}^{\delta_{s}\sqrt{\hat{s}}/2}dE_{5}E_{5}^{1-2\epsilon}\int_{0}^{\pi}\sin^{1-2\epsilon}\varphi_{1}d\varphi_{1}\int_{0}^{\pi}\sin^{-2\epsilon}\varphi_{2}d\varphi_{2}.
\end{equation}
After performing the integrations we arrive at a form where IR
singularities are explicit
\begin{equation}
d\hat{\sigma}^{S}=d\hat{\sigma}^{B}\frac{\alpha_{s}}{2\pi}C(\epsilon)\Big(\frac{A_{2}^{S}}{\epsilon^{2}}+\frac{A_{1}^{S}}{\epsilon}+A_{0}^{S}\Big),
\end{equation}
with
\begin{equation}
A_{2}^{S}=2C_{F},\qquad A_{1}^{S}=-4C_{F}\ln\delta_{s},\qquad A_{0}^{S}=4C_{F}\ln^{2}\delta_{s}.
\end{equation}

\subsubsection{Hard collinear region}

In the hard collinear region, the factorization theorem~\cite{fac-theorem1,fac-theorem2} states that the squared amplitude can be factorized into the squared Born amplitude
multiplied by the unregulated Altarelli-Parisi splitting function
as long as the matrix element is calculated under
the collinear limit of kinematic configuration.
\begin{equation}
\overline{|\mathcal{M}^{R}(b\bar{b}\rightarrow\gamma A^{0}+g)|^{2}}\overset{coll.}{\longrightarrow}(4\pi\alpha_{s}\mu_{R}^{2\epsilon})\overline{|\mathcal{M}^{B}(b^{\prime}\bar{b}\rightarrow\gamma A^{0};\hat{s}^{\prime}=z\hat{s})|^{2}}\frac{-2P_{b^{\prime}b}(z,\epsilon)}{z(p_{1}-p_{5})^{2}}.
\end{equation}
Moreover, the phase space can also be factorized in the collinear limit,
\begin{equation}
dPS^{(3)}(b\bar{b}\rightarrow\gamma A^{0}+g)\overset{coll.}{\longrightarrow}dPS^{(2)}(b^{\prime}\bar{b}\rightarrow\gamma A^{0})\frac{(4\pi)^{\epsilon}}{16\pi^{2}\Gamma(1-\epsilon)}dzdt_{15}\big[-(1-z)t_{15}\big]^{-\epsilon},
\end{equation}
with $t_{15}=(p_{1}-p_{5})^{2}$. After convoluting with the PDFs we obtain an expression
for the inclusive cross section where collinear singularities are explicit
in terms of $1/\epsilon$ poles~\cite{PS-slicing}
\begin{multline}
d\sigma^{HC}_{b-{\rm{splitting}}}=d\hat{\sigma}^{B}(b\bar{b}\rightarrow\gamma A^{0})\Big[G_{b/p}\big(\frac{x_{1}}{z}\big)G_{\bar{b}/p}(x_{2})+(x_{1}\leftrightarrow x_{2})\Big]\\
\times\frac{\alpha_{s}}{2\pi}C(\epsilon)\Big(\frac{1}{-\epsilon}\Big)\delta_{c}^{-\epsilon}P_{bb}(z,\epsilon)\frac{dz}{z}\Big(\frac{1-z}{z}\Big)^{-\epsilon}dx_{1}dx_{2}.
\end{multline}
A similar term which gives exactly the same contribution is also present
to account for initial-state anti-quark splitting. So the complete
collinear piece is
\begin{multline}
d\sigma^{HC}=\Big[G_{b/p}\big(\frac{x_{1}}{z}\big)G_{\bar{b}/p}(x_{2})+G_{\bar{b}/p}\big(\frac{x_{1}}{z}\big)G_{b/p}(x_{2})+(x_{1}\leftrightarrow x_{2})\Big]d\hat{\sigma}^{B}(b\bar{b}\rightarrow\gamma A^{0})\\
\times\frac{\alpha_{s}}{2\pi}C(\epsilon)\Big(\frac{1}{-\epsilon}\Big)\delta_{c}^{-\epsilon}P_{bb}(z,\epsilon)\frac{dz}{z}\Big(\frac{1-z}{z}\Big)^{-\epsilon}dx_{1}dx_{2},
\end{multline}
where the unregulated Altarelli-Parisi splitting functions are written
explicitly as
\begin{equation}
P_{bb}(z,\epsilon)=P_{\bar{b}\bar{b}}(z,\epsilon)=C_{F}\Big(\frac{1+z^{2}}{1-z}-\epsilon(1-z)\Big).
\end{equation}
where $G_{b(\bar{b})/p}(x)$ is temporarily the bare PDF.
Due to the non-soft constraint we have $x_{1}\leq z\leq1-\delta_{s}$.

\subsection{Massless bottom (anti-)quark emission}

\begin{figure}[H]
\begin{centering}
\includegraphics[scale=0.5]{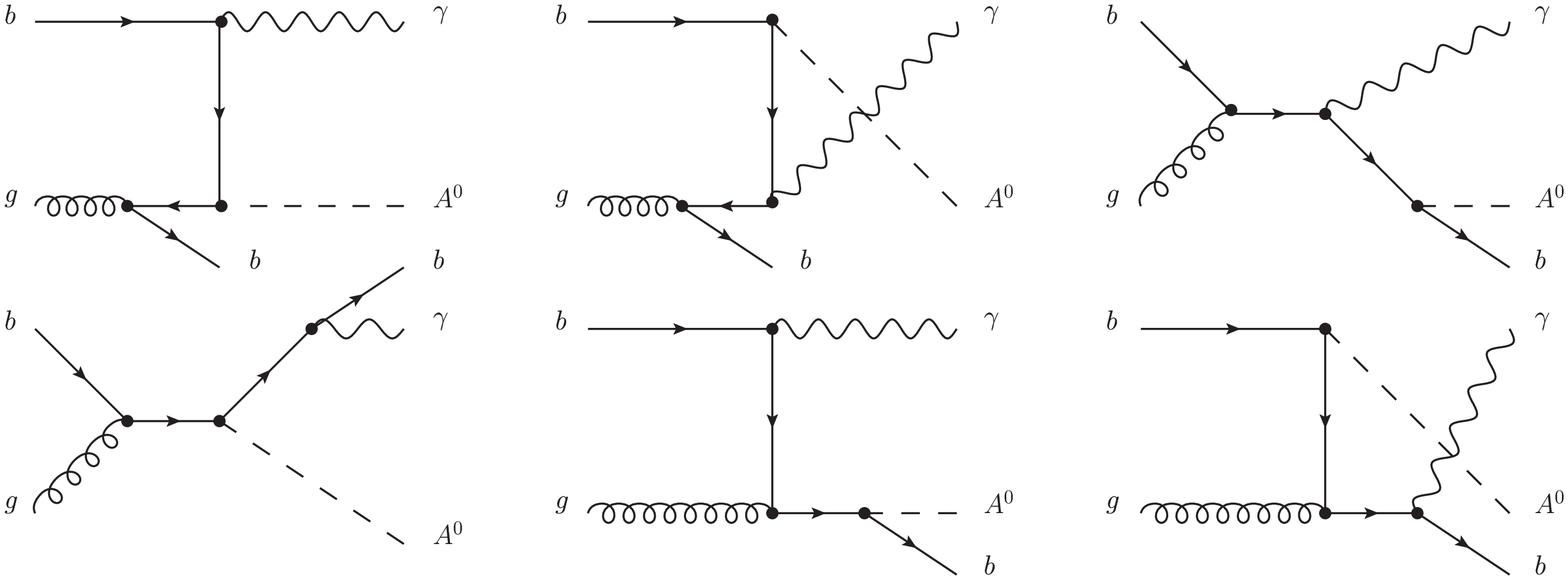}
\par\end{centering}

\caption{\label{fig:bg-initial}Feynman diagrams for massless bottom (anti-)quark
emission}

\end{figure}

At $\mathcal{O}(\alpha_{s})$ of the perturbative expansion $bg$ (or $\bar{b}g$) initial subprocesses
should be taken into consideration, with the relevant Feynman diagrams shown in Fig.\ref{fig:bg-initial}.

The treatment is much the same as to $b\bar{b}$ annihilation except for some differences. First, the radiation of a massless (anti-)quark contains no soft divergence.
Hence there is no need to introduce a soft cutoff, and the 3-body phase
space is divided into a collinear region and a non-collinear region, for the latter numerical calculation is straightforward.
There is also collinear singularity arising from collinear emission of
massless (anti-)quark. The factorization treatment in the previous
subsection applies if we introduce the collinear cutoff $\delta_{c}$
to separate the collinear region and isolate the collinear poles.
Combining the non-collinear piece and collinear piece we obtain the cross
section
\begin{multline}
d\sigma^{add.}=\underset{\alpha=b,\bar{b}}{\sum}d\hat{\sigma}^{\overline{C}}(g\alpha\rightarrow\gamma A^{0}+\alpha)\big[G_{g/p}(x_{1})G_{\alpha/p}(x_{2})+(x_{1}\leftrightarrow x_{2})\big]dx_{1}dx_{2}\\
+d\hat{\sigma}^{B}(b\bar{b}\rightarrow\gamma A^{0})\frac{\alpha_{s}}{2\pi}C(\epsilon)\Big(-\frac{1}{\epsilon}\Big)\delta_{c}^{-\epsilon}\Big[P_{bg}(z,\epsilon)G_{g/p}\Big(\frac{x_{1}}{z}\Big)G_{\bar{b}/p}(x_{2})+P_{\bar{b}g}(z,\epsilon)G_{g/p}\Big(\frac{x_{1}}{z}\Big)G_{b/p}(x_{2})\\
+(x_{1}\leftrightarrow x_{2})\Big]
\times\frac{dz}{z}\Big(\frac{1-z}{z}\Big)^{-\epsilon}dx_{1}dx_{2},
\end{multline}
where the unregulated Altarelli-Parisi splitting functions are written
explicitly as
\begin{equation}
P_{bg}(z,\epsilon)=P_{\bar{b}g}(z,\epsilon)=\frac{3}{8}C_{F}\big(z^{2}+(1-z)^{2}-2z(1-z)\epsilon\big).
\end{equation}
Further collinear singularity can still arise in the configuration
in which the photon is emitted in parallel with the additional
final-state quark. By comparison, such singularity does not exist
for a final-state gluon at next-to-leading order. A criterion for
isolated photon has been suggested in Refs.~\cite{isolated-photon},
which defines an IR-safe cross section decoupled with hadronic
fragmentation and at the same time allows for complete cancelation
of soft gluon divergence. For the case of only one final-state
parton such criterion is equivalent to the kinematic cut
\begin{equation}
p^{j}_{T}<\frac{1-\cos\Delta R_{j\gamma}}{1-\cos\Delta R_{0}}p^{\gamma}_{T},\qquad\mathrm{for}~\Delta R_{j\gamma}<\Delta R_{0},
\end{equation}
where $j$ stands for either the final-state (anti-)quark or the final-state gluon, and $\Delta R_{j\gamma}$ is the cone distance in the rapidity-azimuthal angle plane between the parton and the photon. Throughout our calculation we choose the cone-size parameter $\Delta R_{0}=0.4$.

\subsection{Mass factorization}

Since the real correction and the virtual correction combined are
incomplete to cancel all the divergences, the procedure of mass
factorization is necessary. Generally, the scale-dependent
PDF $G_{\alpha/\beta}(x,\mu_{F})$ under $\overline{MS}$ scheme can
be written following Ref.~\cite{PS-slicing}
\begin{equation}
G_{\alpha/p}(x,\mu_{F})=G_{\alpha/p}(x)+\underset{\beta}{\sum}\Big(-\frac{1}{\epsilon}\Big)\frac{\alpha_{s}}{2\pi}C(\epsilon)\Big(\frac{\mu_{F}^{2}}{\hat{s}}\Big)^{\epsilon}\int_{x}^{1}\frac{dz}{z}P_{\alpha\beta}(z)G_{\beta/p}\Big(\frac{x}{z}\Big).
\end{equation}
The Altarelli-Parisi splitting function in the
above formula is independent of $\epsilon$ which is defined by
\begin{equation}
P_{\alpha\beta}(y,\epsilon)=P_{\alpha\beta}(y)+\epsilon P_{\alpha\beta}^{\prime}(y).
\end{equation}
Thus a collinear counter-term of $\mathcal{O}(\alpha_{s})$ is obtained
from the LO piece and will be used to cancel the collinear divergence.
If we combine the counter-term with the hard collinear pieces, from both
$b\bar{b}$ channel and $bg$ (or $\bar{b}g$) channel, we will find
the remaining collinear piece in the following form,
\begin{multline}
d\sigma^{C}=d\hat{\sigma}^{B}\frac{\alpha_{s}}{2\pi}C(\epsilon)\Big\{\tilde{G}_{b/p}(x_{1},\mu_{F})G_{\bar{b}/p}(x_{2},\mu_{F})+G_{b/p}(x_{1},\mu_{F})\tilde{G}_{\bar{b}/p}(x_{2},\mu_{F})\\
+\underset{\alpha=b,\bar{b}}{\sum}\Big[\frac{A_{1}^{SC}(\alpha\rightarrow\alpha g)}{\epsilon}+A_{0}^{SC}(\alpha\rightarrow\alpha g)\Big]G_{b/p}(x_{1},\mu_{F})G_{\bar{b}/p}(x_{2},\mu_{F})+(x_{1}\leftrightarrow x_{2})\Big\}dx_{1}dx_{2}.
\end{multline}
The summed terms with $A_{1,0}^{SC}$ are a result of an overlap of both soft and collinear phase space regions.
One can explicitly write
\begin{align}
A_{1}^{SC}(b & \rightarrow bg)=A_{1}^{SC}(\bar{b}\rightarrow\bar{b}g)=C_{F}\Big(2\ln\delta_{s}+\frac{3}{2}\Big),\nonumber\\
A_{0}^{SC}(b & \rightarrow bg)=A_{0}^{SC}(\bar{b}\rightarrow\bar{b}g)=C_{F}\Big(2\ln\delta_{s}+\frac{3}{2}\Big)\ln\frac{\hat{s}}{\mu_{F}^{2}},
\end{align}
and the tilded $G$ functions
\begin{equation}
\tilde{G}_{\alpha/p}(x,\mu_{F})=\underset{\beta}{\sum}\int_{x}^{1-\delta_{s}\delta_{\alpha\beta}}\frac{dy}{y}\tilde{P}_{\alpha\beta}(y)G_{\beta/p}\Big(\frac{x}{y},\mu_{F}\Big),
\end{equation}
with
\begin{equation}
\tilde{P}_{\alpha\beta}(y)=P_{\alpha\beta}(y)\ln\Big(\delta_{c}\frac{1-y}{y}\frac{\hat{s}}{\mu_{F}^{2}}\Big)-P_{\alpha\beta}^{\prime}(y).
\end{equation}
Now we can confirm that all the divergences have been canceled, since
\begin{align}
2A_{2}^{V}+A_{2}^{S} & =0,\nonumber\\
2A_{1}^{V}+A_{1}^{S}+\underset{\alpha=b,\bar{b}}{\sum}A_{1}^{SC}(\alpha\rightarrow\alpha g) & =0.
\end{align}
Putting together all pieces, we find a finite result of the NLO
QCD total cross section for $pp\rightarrow\gamma A^{0}+X$
\begin{multline}
\sigma^{NLO}=\int\Bigg\{dx_{1}dx_{2}\Big[G_{b/p}(x_{1},\mu_{F})G_{\bar{b}/p}(x_{2},\mu_{F})+(x_{1}\leftrightarrow x_{2})\Big](\hat{\sigma}^{B}+\hat{\sigma}^{V}+\hat{\sigma}^{S}+\hat{\sigma}^{\overline{HC}})+\hat{\sigma}^{C}\Bigg\}\\
+\underset{\alpha=b,\bar{b}}{\sum}\int dx_{1}dx_{2}\Big[G_{\alpha/p}(x_{1},\mu_{F})G_{g/p}(x_{2},\mu_{F})+(x_{1}\leftrightarrow x_{2})\Big]\hat{\sigma}^{\overline{C}}(\alpha g\rightarrow\gamma A^{0}+\alpha).
\end{multline}
We see that the total cross section depends on two undetermined
scales: the renormalization scale $\mu_{R}$ and the factorization
scale $\mu_{F}$.

\section{MONTE CARLO SIMULATION\label{sec:SIMULATION}}

At the LHC, the leptonic decay mode $A^0\rightarrow\tau^{+}\tau^{-}$ will be the most promising signature in the search of $A^0$. For a moderate Higgs mass, the branch ratio $\Gamma_{A^0\rightarrow\tau^{+}\tau^{-}}$ is around $10\%$, but the QCD background is much smaller compared with that in the decay mode $A^0\rightarrow b\bar{b}$. Therefore we also conduct a Monte Carlo simulation study of the $\tau^{+}\tau^{-}+\gamma$ signature against the dominant irreducible SM background, namely the off-shell production of gauge bosons $q\bar{q}\rightarrow\gamma Z^{*}/\gamma\gamma^{*}\rightarrow\gamma\tau^{+}\tau^{-}$. For the calculation of the background we use the package CompHep v4.5.1~\cite{CompHep}.

We impose the transverse momentum cuts $p_{T}^{\gamma}>30$~GeV, $p_{T}^{\tau}>20$~GeV, and the pseudo-rapidity cuts $|\eta^{\gamma,\tau}|<2.5$ for the photon and the tau leptons. We require the distance $\Delta R_{\gamma\tau}>0.7$, $\Delta R_{\tau\tau}>0.7$ to ensure well-separated final states. To reconstruct the on-shell Higgs boson $A^0$, we also demand that the tau pair invariant mass is within the window $[0.9m_A, 1.1m_{A}]$. Besides, an additional cut on the azimuthal angles $\Delta\phi_{\tau\tau}<2.9$ is also imposed on the tau lepton pair, which is very effective at suppressing the false signature arising from a high-$p_{T}$ photon radiated from one of the tau leptons. After all the above kinematic cuts are applied, the SM background cross section can be reduced by 3 orders of magnitude. All these cuts, summarized in Tab.~\ref{tab:cut}, are in accord with Ref.~\cite{tree-level} in order to compare our results with theirs.

\begin{table*}
\begin{center}
\scalebox{1}[0.9]
{\begin{tabular}{c}
  \hline
  \hline
  \ Kinematic cuts\
  \\
  \hline
  $p_{T}^{\gamma}>30$~GeV,\ $p_{T}^{\tau}>20$~GeV
  \\
  $|\eta^{\gamma,\tau}|<2.5$
  \\
  $\Delta R_{\gamma\tau}>0.7$,\ $\Delta R_{\tau\tau}>0.7$
  \\
  $0.9m_{A}<M_{\tau\tau}<1.1m_{A}$
  \\
  $\Delta\phi_{\tau\tau}<2.9$
  \\
  \hline
  \hline
\end{tabular}}
\end{center}
\caption{Kinematic cuts imposed in the Monte Carlo simulation} \label{tab:cut}
\end{table*}

\section{NUMERICAL RESULTS\label{sec:NUMERICAL}}

This section is arranged as follows. First, we present the numerical results for the complete NLO QCD corrected cross sections to $A^0\gamma$ associated production. Then we present simulation results of the $\tau^{+}\tau^{-}+\gamma$ signature under various kinematic cuts, for both the integrated cross section and differential cross sections. It is worth to mention that in our results $H^0-A^{0}$ degeneracy is not assumed. For large $m_{A}$ and large $\tan\beta$, such degeneracy doubles the cross section.

\subsection{NLO total cross section calculations}

In this section, we present the results of the inclusive total cross section for $pp\rightarrow\gamma A^{0}+X$
at the LHC with total colliding energy $\sqrt{s}=14$~TeV. Throughout our calculations CTEQ6L1 parton structure functions are used for LO cross sections and CTEQ6M used for the NLO ones. We impose the photon transverse momentum cut $p^{\gamma}_{T}>30$~GeV and pseudo-rapidity cut $|\eta^{\gamma}|<2.5$.
We choose the following SM input parameters~\cite{pdg}
\begin{multline}
m_{t}=172.4~{\rm{GeV}},\quad G_{F}=1.16637\times10^{-5}~{\rm{GeV}}^{-2},\quad m_{W}=80.398~{\rm{GeV}},\quad m_{Z}=91.1876~{\rm{GeV}},\\
\quad\alpha_{s}(m_{Z})=0.1176,\quad m_{b}^{pole}=4.68~{\rm{GeV}},\quad m_{b}(m_{b}^{pole})=4.20~{\rm{GeV}},\quad\alpha_{em}(m_{W})=1/128.
\end{multline}
Both the strong coupling $\alpha_{s}$ and the running bottom quark mass~\cite{running-mass} are evolved up to two loops in QCD
\begin{equation}
m_{b}(\mu_{R})=U_{6}(\mu_{R},m_{t})U_{5}(m_{t},m_{b}^{pole})m_{b}(m_{b}^{pole}),
\end{equation}
where the evolution factor $U_{f}$ is given by
\begin{align}
&U_{f}(\mu_{2},\mu_{1})=\Big(\frac{\alpha_{s}(\mu_{2})}{\alpha_{s}(\mu_{1})}\Big)^{d^{(f)}}\Big[1+\frac{\alpha_{s}(\mu_{1})-\alpha_{s}(\mu_{2})}{4\pi}J^{(f)}\Big],\\
&d^{(f)}=\frac{12}{33-2f},\qquad J^{(f)}=-\frac{8982-504f+40f^{2}}{3(33-2f)^{2}},
\end{align}
and $f$ denotes the number of active quark flavors.

In the large $\tan\beta$ scenario, perturbative calculation is improved by resuming the $\tan\beta$-enhanced
threshold SUSY QCD corrections~\cite{running-mass}. It is equivalent to make the following replacement
for the tree-level bottom quark running mass
\begin{equation}
m_{b}(\mu_{R})\rightarrow\frac{m_{b}(\mu_{R})}{1+\Delta_{b}(\mu_{SUSY})},\qquad\Delta_{b}(\mu_{SUSY})=\frac{\alpha_{s}(\mu_{SUSY})}{2\pi}C_{F}m_{\tilde{g}}\mu I(m_{\tilde{b}_{1}},m_{\tilde{b}_{2}},m_{\tilde{g}})\tan\beta,
\end{equation}
where the auxiliary function is defined by
\begin{equation}
I(a,b,c)=-\frac{1}{(a^{2}-b^{2})(b^{2}-c^{2})(c^{2}-a^{2})}\Big(a^{2}b^{2}\ln\frac{a^{2}}{b^{2}}+b^{2}c^{2}\ln\frac{b^{2}}{c^{2}}+c^{2}a^{2}\ln\frac{c^{2}}{a^{2}}\Big),
\end{equation}
To avoid double-counting, an additional finite counter-term for the bottom quark mass should
be introduced
\begin{equation}
\frac{\delta\tilde{m}_{b}}{m_{b}}=\Delta_{b}\Big(1+\frac{1}{\tan^{2}\beta}\Big).
\end{equation}

For the SUSY QCD contribution the package SPheno v2.2.2 is used
to calculate all the parameters in the MSSM~\cite{spheno}. We choose the
minimal supergravity scenario (mSUGRA) in which various MSSM parameters
are constrained by only five free input parameters at
the grand unification scale: $m_{1/2}$, $m_{0}$, $A_{0}$, $\tan\beta$
and the sign of $\mu$. The first three parameters $m_{1/2}$, $m_{0}$,
$A_{0}$ are, respectively, the universal gaugino mass, the universal scalar mass,
and the trilinear soft breaking parameter of the superpotential~\cite{mSUGRA}. We
fix $m_{1/2}=200$~GeV, $A_{0}=0$ while $\tan\beta$ and the
sign of $\mu$ are left as free parameters. The desired value of $m_{A}$
is obtained by tuning $m_{0}$. Unless specified, the factorization scale $\mu_{F}$ and the renormalization scale $\mu_{R}$ are always set equal at $\mu_{F}=\mu_{R}=\mu_{0}=m_{A}/2$.
Besides, a third scale, the SUSY scale
$\mu_{SUSY}$ which comes into effect by threshold SUSY QCD resummation to
the bottom quark Yukawa coupling, is chosen to be $\mu_{SUSY}=2$~TeV.

\begin{figure}[H]
\begin{centering}
\includegraphics[scale=0.9]{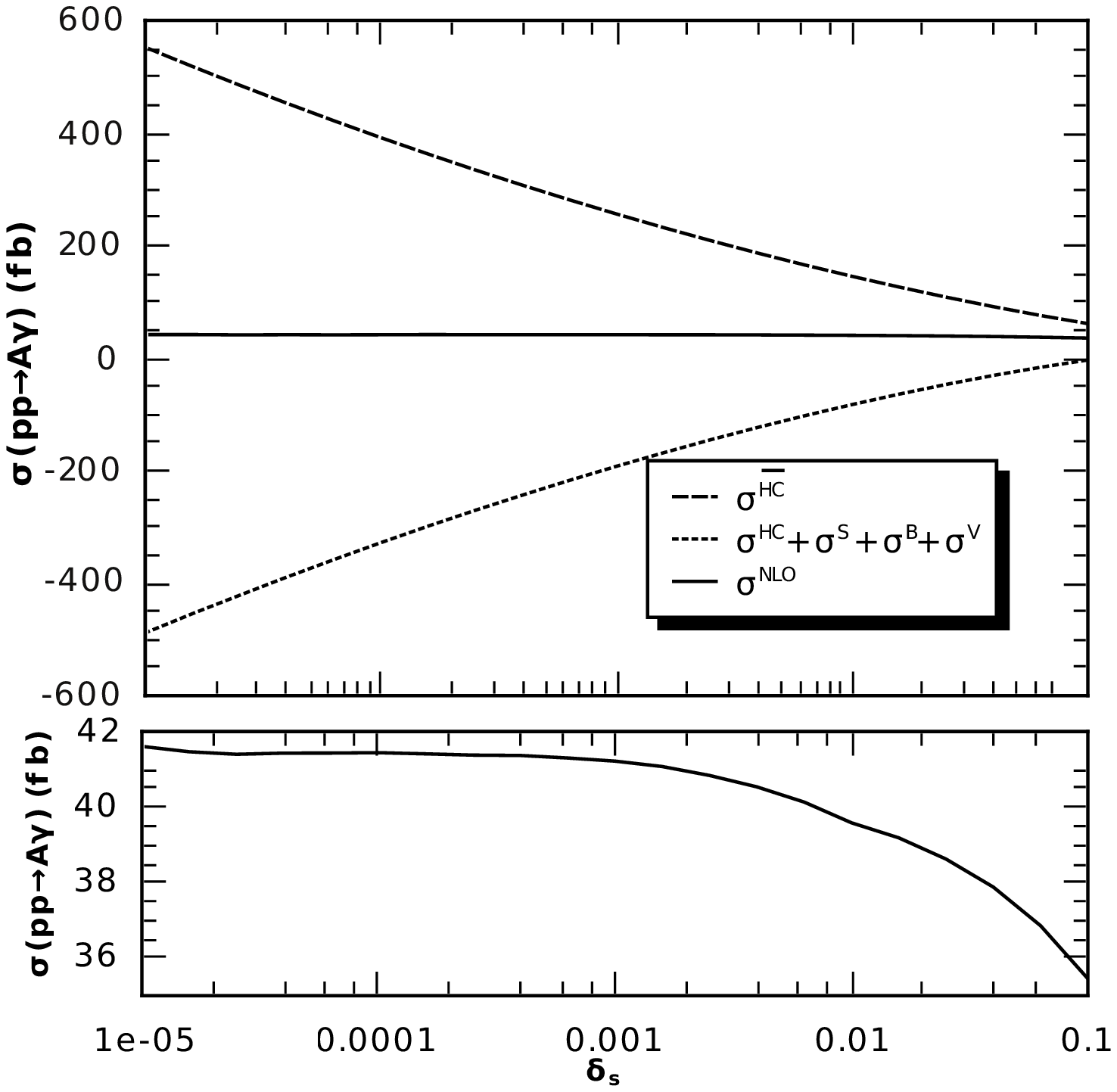}
\par\end{centering}

\caption{\label{fig:cutoff-dependence} Inclusive total cross sections for $pp\rightarrow A^{0}\gamma+X$ at the LHC as a function of $\delta_{s}$ in the phase space slicing treatment. Non-collinear real correction, collinear correction, soft and virtual corrections are also shown separately. The collinear cutoff is chosen to be $\delta_{c}=\delta_{s}/50$.}

\end{figure}

\begin{figure}[H]
\begin{centering}
\includegraphics[scale=0.8]{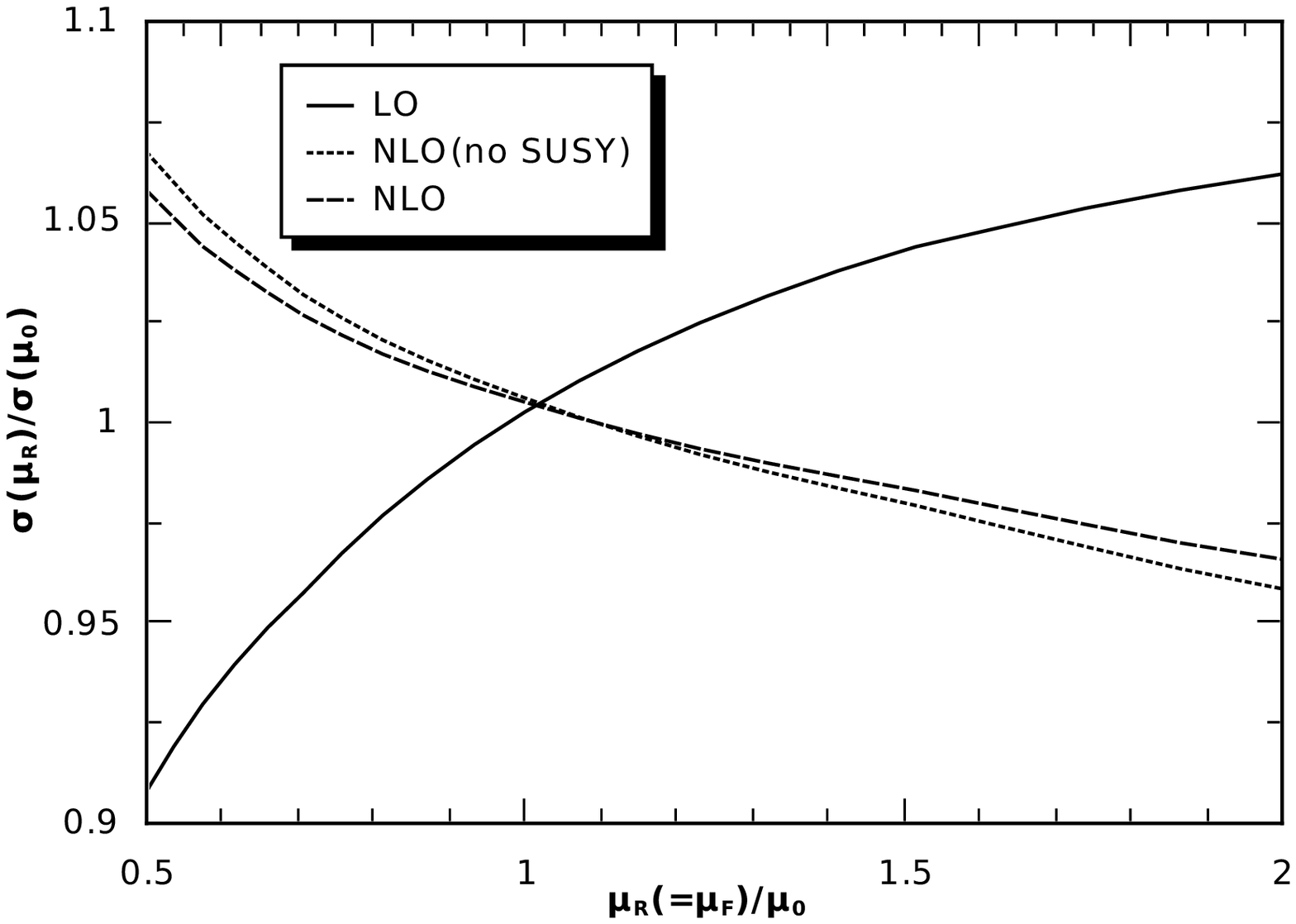}
\par\end{centering}

\caption{\label{fig:scale-dependence} Dependence of inclusive total cross section for $pp\rightarrow A^{0}\gamma+X$ at the LHC on the factorization scale and the renormalization scale assuming $\mu_{R}=\mu_{F}$.}

\end{figure}

In Fig.\ref{fig:cutoff-dependence}, the NLO total cross section is
plotted against $\delta_{s}$ and $\delta_{c}$ over a wide range of
variation at the SUSY benchmark point SPS 4~\cite{SPS}. For the NLO
corrections, the real/hard correction depends on $\delta_{s}$ and
$\delta_{c}$ , the virtual and soft gluon pieces combined depends
only on $\delta_{s}$, and the hard collinear part depends only on
$\delta_{c}$. However, when all pieces are added together, the
dependence on $\delta_{s}$ and $\delta_{c}$ is canceled out as long
as sufficiently small values of $\delta_{s}$ and $\delta_{c}$ are
chosen. From Fig. \ref{fig:scale-dependence}, in which the SPS 4
benchmark point is also chosen, we can see that the complete NLO QCD
corrections improve the scale dependence as compared to the LO
results for $m_{A}/4<\mu_{0}<m_{A}$. In addition, the SUSY QCD
correction is found to further reduce the scale uncertainty even
though it is much smaller than the SM QCD correction.

\begin{figure}[H]
\begin{centering}
\includegraphics[scale=0.5]{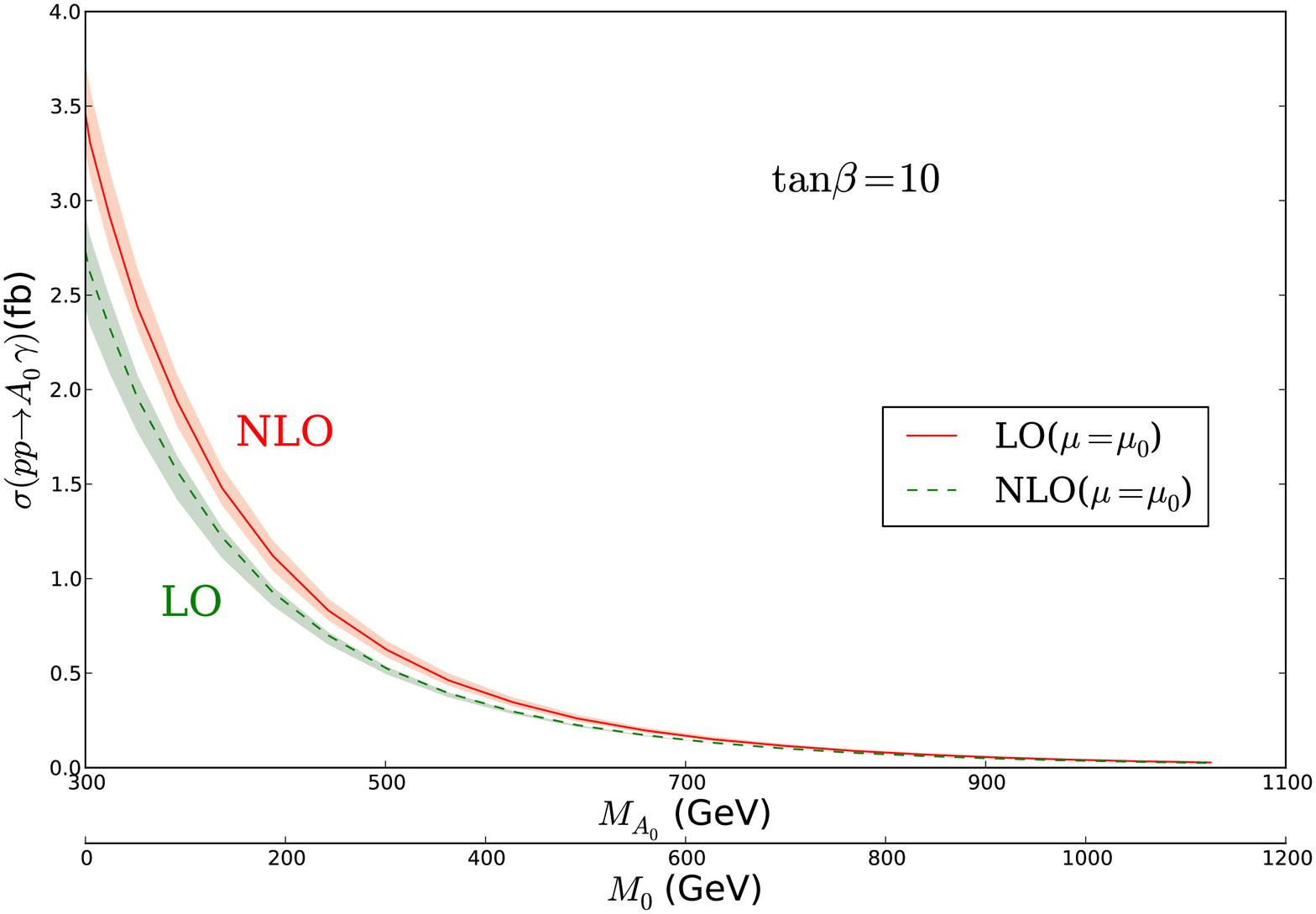}
\par\end{centering}

\caption{\label{fig:ma0tb10} The inclusive total cross sections for
$pp\rightarrow A^{0}\gamma+X$ at the LHC as a function of $m_{A^0}$,
with $\tan\beta=10$. The bands are obtained by varying the
renormalization and factorization scale between
$\mu_R(=\mu_F)=\mu_0/2$ and $\mu_R(=\mu_F)=2\mu_0$ .}

\end{figure}

\begin{figure}[H]
\begin{centering}
\includegraphics[scale=0.5]{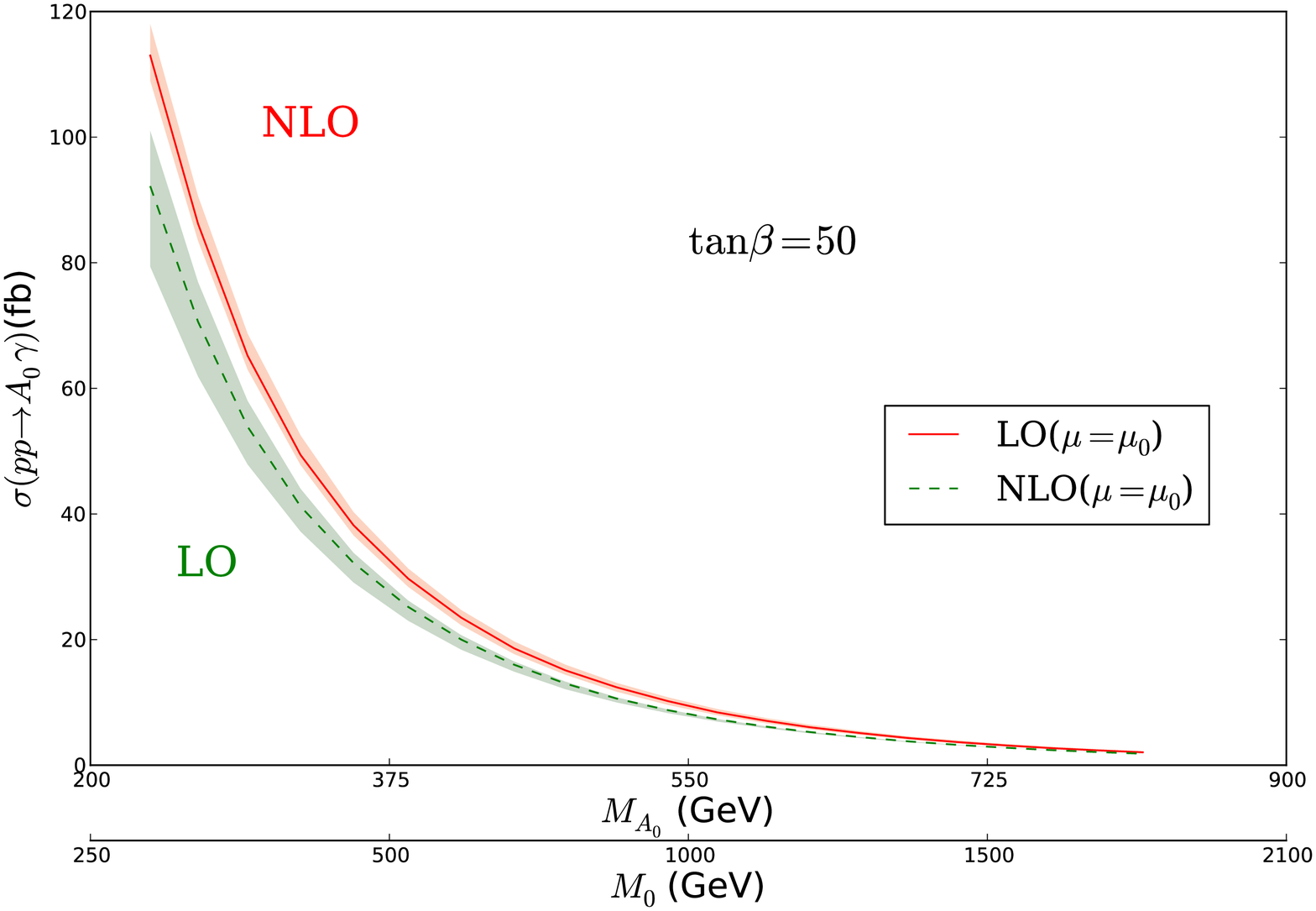}
\par\end{centering}

\caption{\label{fig:ma0tb50} The inclusive total cross sections for
$pp\rightarrow A^{0}\gamma+X$ at the LHC as a function of $m_{A^0}$,
with $\tan\beta=50$. The bands are obtained by varying the
renormalization and factorization scale between
$\mu_R(=\mu_F)=\mu_0/2$ and $\mu_R(=\mu_F)=2\mu_0$.}

\end{figure}

\begin{figure}[H]
\begin{centering}
\includegraphics[scale=0.5]{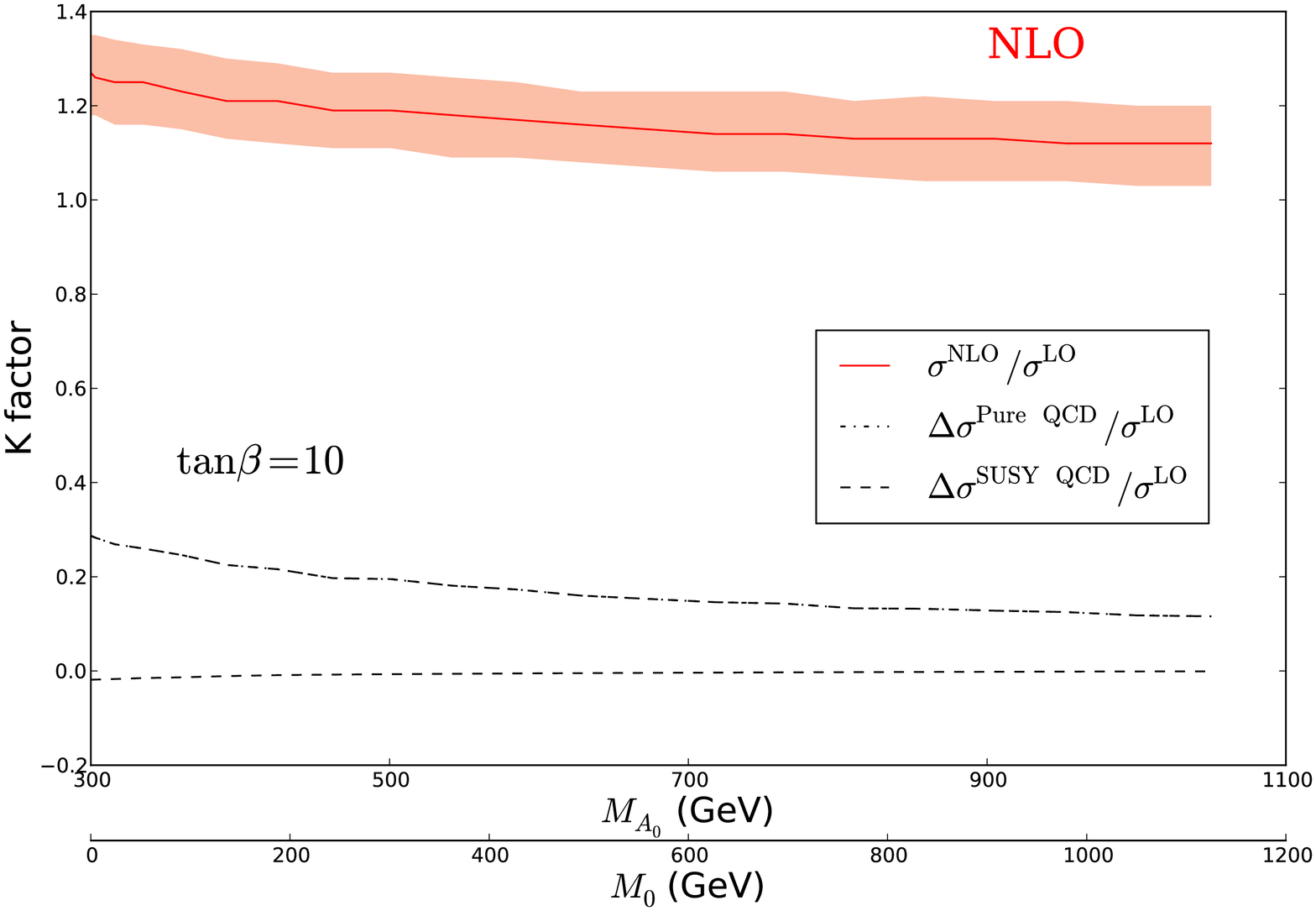}
\par\end{centering}

\caption{\label{fig:kftb10} K-factors for $pp\rightarrow
A^{0}\gamma+X$ at the LHC with $\tan\beta=10$. The band is obtained
by varing the scale in the NLO calculations between
$\mu_R(=\mu_F)=\mu_0/2$ and $\mu_R(=\mu_F)=2\mu_0$.}

\end{figure}

\begin{figure}[H]
\begin{centering}
\includegraphics[scale=0.5]{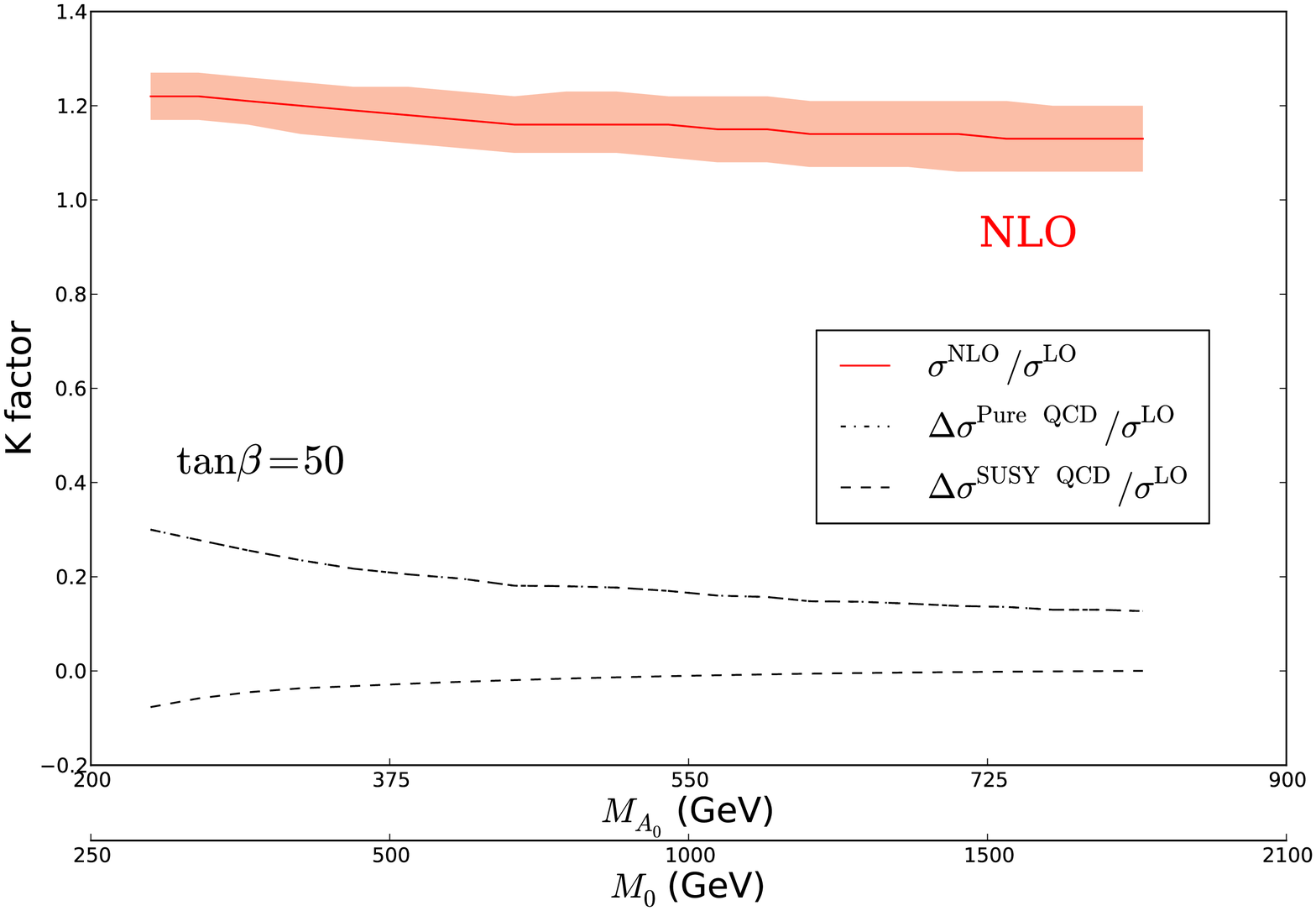}
\par\end{centering}

\caption{\label{fig:kftb50} K-factors for $pp\rightarrow
A^{0}\gamma+X$ at the LHC with $\tan\beta=50$. The band is obtained
by varing the scale in the NLO calculations between
$\mu_R(=\mu_F)=\mu_0/2$ and $\mu_R(=\mu_F)=2\mu_0$.}

\end{figure}

In Fig.\ref{fig:ma0tb10} and Fig.\ref{fig:ma0tb50}, we plot the
total cross sections with scale uncertainties for the inclusive
$pp\rightarrow\gamma A^{0}+X$ production as functions of the Higgs
boson mass $m_{A}$. A positive MSSM soft breaking parameter $\mu$,
which is favored by the measurement of
$(g-2)_{\mu}$~\cite{gminustwo}, is of particular interest. However,
in the $\mu>0$ scenario the mass of $A^{0}$ can not be smaller than
$200$~GeV.  Assuming $\tan\beta=50$, the total cross section
decreases rapidly as the Higgs boson becomes heavier, from
$60\sim70$~fb for relatively light Higgs boson mass $m_{A}=300$~GeV
to a mere $15$~fb for much heavier Higgs boson mass $m_{A}=500$~GeV.
For the case of $\tan\beta=10$ the total cross section is an order
smaller. The NLO corrections efficiently reduce the total scale
dependence of the cross sections in the light Higgs boson mass
region but not the heavy mass region. This is because that the
factorization and renormalization scale dependence cancels exactly
in the heavy mass region at the LO. And we have checked that the
factorization and renormalization scale dependence is indeed
improved seperately.
 Also in Fig.\ref{fig:kftb10} and Fig.\ref{fig:kftb50},
K-factor as a function of the Higgs boson mass $m_{A}$ is plotted to
show how much the NLO QCD corrections can modify the LO prediction.
Taking the case of $\tan\beta=50$ for example, QCD corrections from
the pure SM contributions typically increase the total cross section
by around $22\sim16\%$ for $300$~GeV$\leq m_{A}\leq500$~GeV. The
SUSY QCD corrections can suppress the cross section by as much as
$12\%$ for light Higgs mass $m_A=200$~GeV. Nevertheless, the
suppression drops to less than $2\%$ in magnitude for heavy Higgs
mass $m_A=500$~GeV. The scale uncertainties of the NLO total cross
sections range from 10\% to 20\% of the LO total cross sections with
the varying of $m_{A}$ and $\tan\beta$ as can be seen from
Fig.\ref{fig:kftb10} and Fig.\ref{fig:kftb50}.

\subsection{Simulation results}

In Tab.\ref{tab:signal}, we present the results of the integrated signal cross section including the LO results, the NLO results without the SUSY QCD corrections, and the complete NLO results. For the mSUGRA input parameters, we fix $m_{1/2}=200$~GeV, $A_{0}=0$, $\tan\beta=50$ and $\mu>0$, and tune $m_{0}$ to obtain Higgs mass $m_{A}=200$~GeV, $300$~GeV, $500$~GeV. For the heavier Higgs mass cases, we choose transverse momentum cut, $p_{T}^{\gamma}>40$~GeV, $50$~GeV for $m_{A}=300$~GeV, $500$~GeV, respectively. Other cuts are the same as what has been mentioned in Sec.~\ref{sec:SIMULATION}. Moreover, an integrated luminosity of $100$~fb$^{-1}$ and a $\tau$-pair detection efficiency $\epsilon_{\tau\tau}=0.2$ are assumed to evaluate the signal significance $\mathcal{S}=N(S)/\sqrt{N(B)}$.

\begin{table*}
\begin{center}
\scalebox{1}[0.9]
{\begin{tabular}{c|c|c|c|c}
  \hline
  \hline
   &Background&LO&NLO (no SUSY)&NLO  \\
  $m_{A}$~[GeV]&~$\sigma_{B}$~[fb]&$\sigma_{S}$~[fb]\quad$\mathcal{S}$&$\sigma_{S}$~[fb]\quad$\mathcal{S}$&$\sigma_{S}$~[fb]\quad$\mathcal{S}$~\\
  \hline
  200 & 3.44 & 8.38\qquad20.2 & 10.8\qquad26.0 & 9.84\qquad23.7\\
  300 & 1.12 & 1.91\qquad8.05 & 2.39\qquad10.0 & 2.30\qquad9.71\\
  500 & 0.270 & 0.287\;\quad2.47 & 0.354\;\quad3.05 & 0.349\;\quad3.00\\
  \hline
  \hline
\end{tabular}}
\end{center}
\caption{\label{tab:signal} Signal cross section $\sigma_{S}$, background cross section $\sigma_{B}$ and significance $\mathcal{S}$ for the associated production $pp\rightarrow A^{0}\gamma\rightarrow\tau^{+}\tau^{-}\gamma$ at the LHC. We set the mSUGRA input parameters $m_{1/2}=200$~GeV, $A_{0}=0$, $\tan\beta=50$ and $\mu>0$.} \label{t1}
\end{table*}

\begin{figure}[H]
\begin{centering}
\includegraphics[scale=0.8]{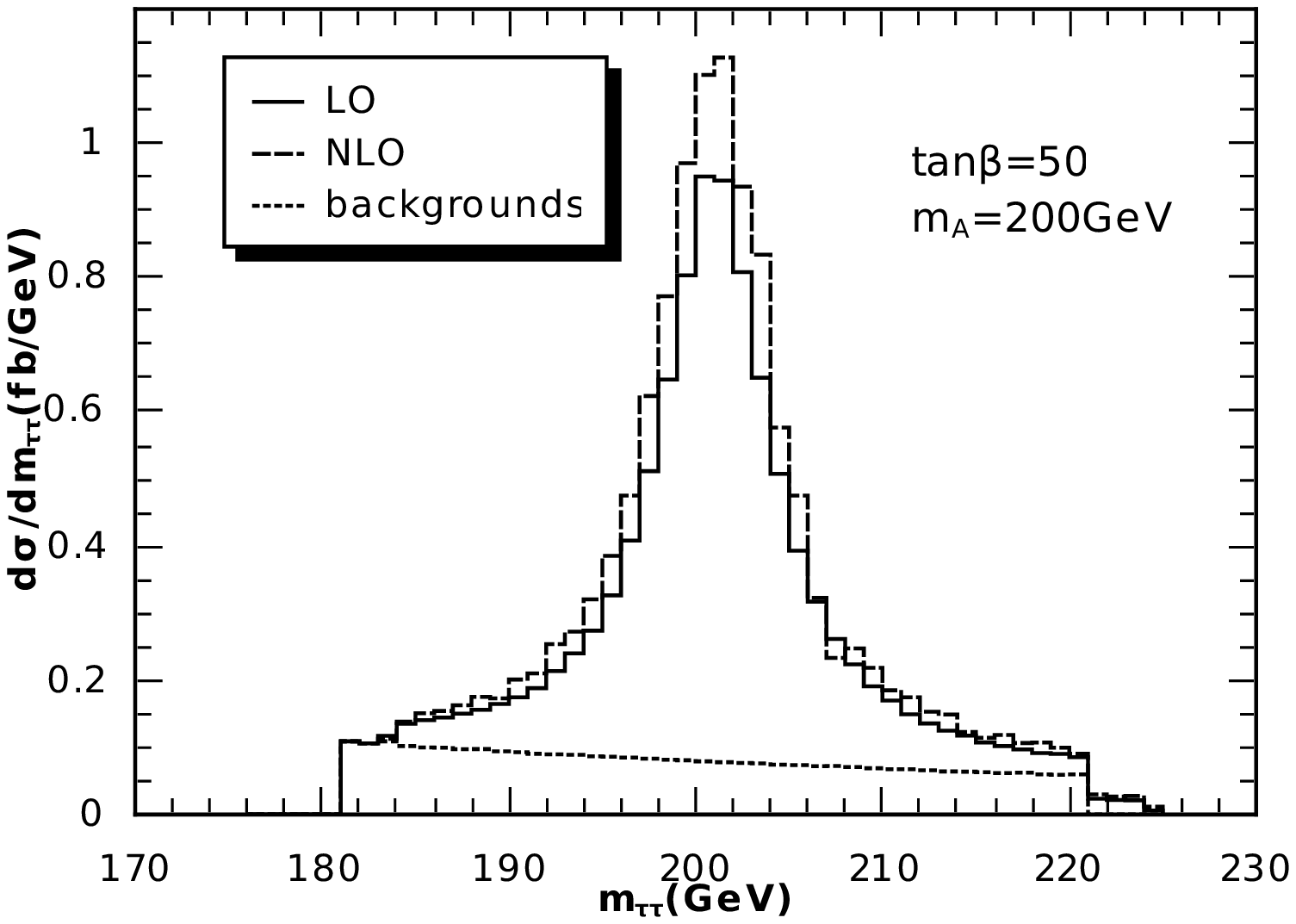}
\par\end{centering}

\caption{\label{fig:ttinvm} Final state $\tau\tau$ invariant mass distribution for $pp\rightarrow A^{0}\gamma+X\rightarrow\tau^{+}\tau^{-}\gamma+X$ at the LHC compared with the background.}

\end{figure}

\begin{figure}[H]
\begin{centering}
\includegraphics[scale=0.8]{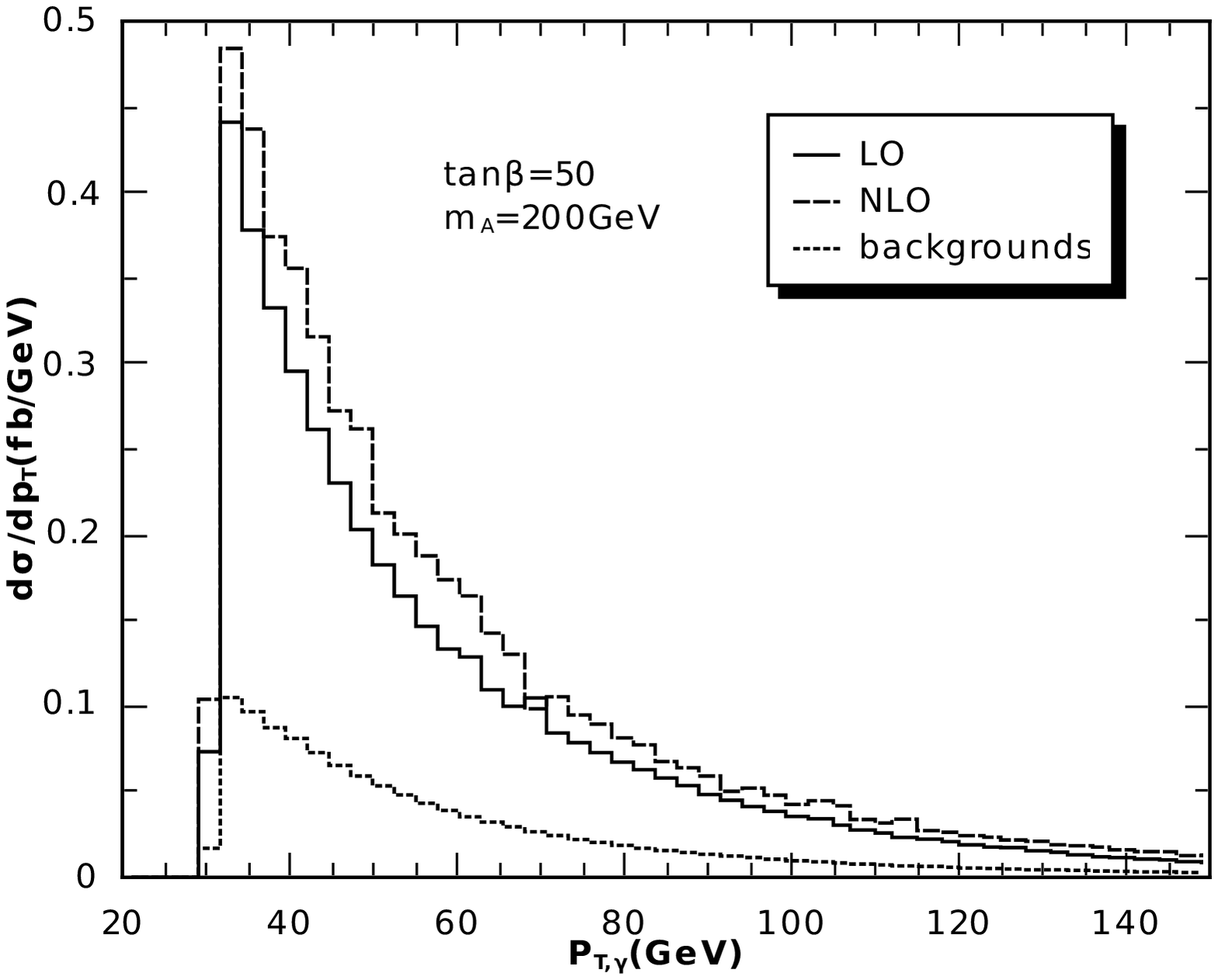}
\par\end{centering}

\caption{\label{fig:ptdistribution} The photon transverse momentum distribution for $pp\rightarrow A^{0}\gamma+X\rightarrow\tau^{+}\tau^{-}\gamma+X$ at the LHC as compared with the background.}

\end{figure}

\begin{figure}[H]
\begin{centering}
\includegraphics[scale=0.8]{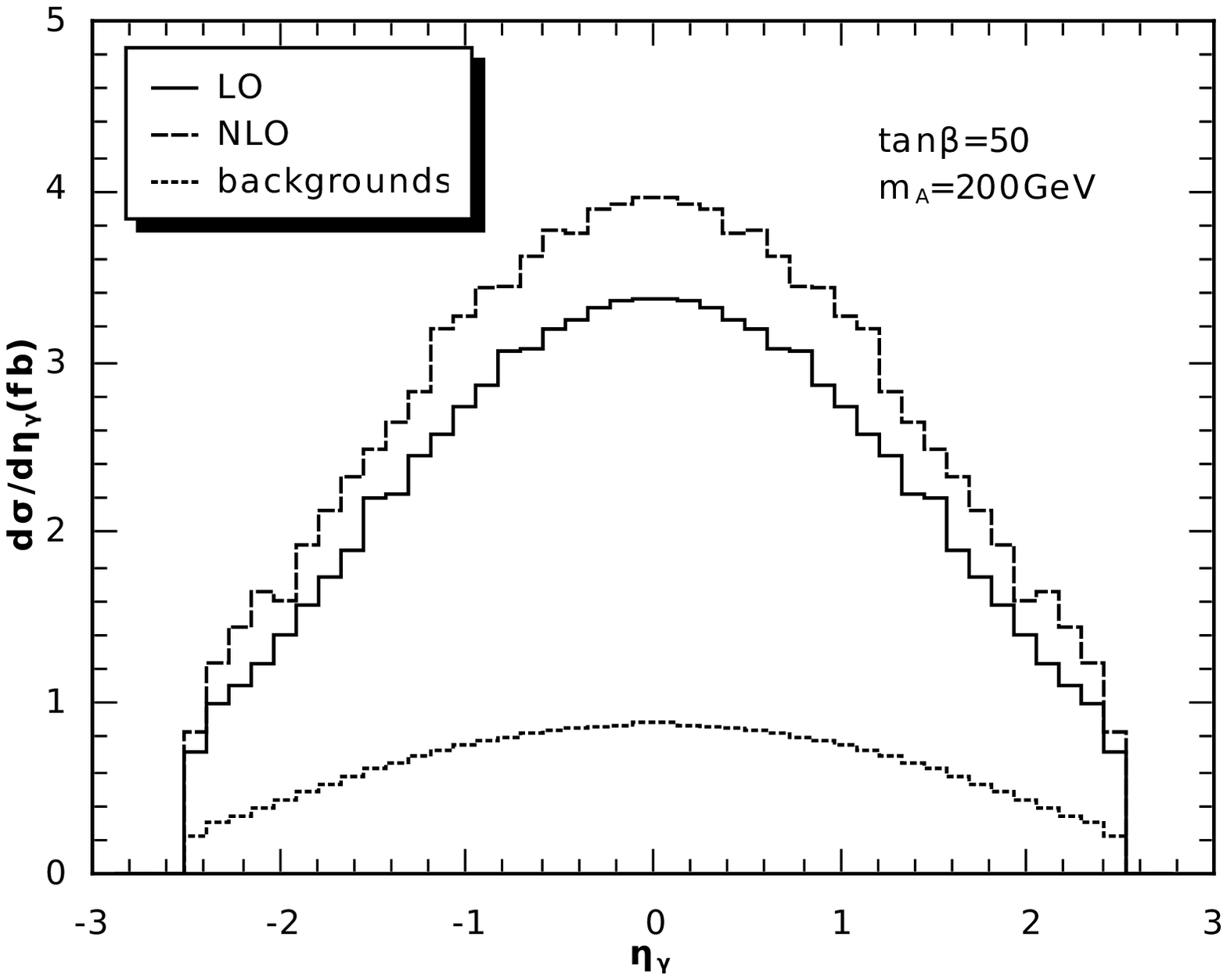}
\par\end{centering}

\caption{\label{fig:etadistribution} Photon transverse momentum distribution for $pp\rightarrow A^{0}\gamma+X\rightarrow\tau^{+}\tau^{-}\gamma+X$ at the LHC as compared with the background.}

\end{figure}

For the case of $m_{A}=200$~GeV in which a relatively large signal cross section and a high significance can be obtained, we investigate the NLO QCD effects more closely by studying various differential cross sections. Fig.\ref{fig:ttinvm} shows the invariant mass distribution $d\sigma/dM_{\tau\tau}$ of the tau lepton pair. With the central region significantly enhanced by the NLO corrections, the mass peak for $A^{0}$ is clearly seen above the background. Fig.\ref{fig:ptdistribution} shows the photon transverse momentum distribution $d\sigma/dp_{T}^{\gamma}$. The NLO QCD effects can enhance the LO results by as much as $13\%$, depending on the specific value of $p_{T}^{\gamma}$. Nevertheless, no significant distortion of the curve is found. In Fig.\ref{fig:etadistribution}, we present the photon pseudo-rapidity distribution $d\sigma/d\eta^{\gamma}$ together with the background. The NLO effects lead to moderate enhancement of the distribution, but do not change the shape of the curve either. Analysis of these differential cross sections shows that the NLO QCD corrections generally enhance the signature.

\section{CONCLUSIONS}
In conclusion, we have investigated the complete NLO QCD corrections
to the inclusive total cross sections of $A^{0}\gamma$ associated
production at the LHC in the MSSM. Our results show that the NLO
corrections can enhance the total cross sections by $25\%\sim15\%$
for Higgs mass $200$~GeV~$<m_{A}<~500$~GeV and $\tan\beta=50$. The
SUSY QCD correction is negative and significant for light Higgs mass
$m_{A}=200$~GeV, but is negligible for heavy Higgs mass
$m_{A}=500$~GeV. The NLO corrections generally reduce the dependence
of the total cross sections on the renormalization/factorization
scale. Assuming a normal luminosity of $100$~fb$^{-1}$, we simulated
the $\tau^{+}\tau^{-}+\gamma$ signature including the complete NLO
QCD effects at the LHC, and found an observable signature above the
SM background with a high signal significance in some regions of the
MSSM parameter space allowed by the current experiments. Thus it can
be expected that the LHC has the potential to discover a CP-odd
Higgs boson with a mass of $200$~GeV$\sim300$~GeV via the photon
associated
   production channel for large $\tan\beta$.

\begin{acknowledgements}
This work was supported in part by the National Natural Science Foundation of China, under Grants No. 10721063, No. 10975004 and No. 10635030.
\end{acknowledgements}

\end{document}